\mag\magstep1

\documentclass{article}

\usepackage{graphicx}
\usepackage{natbib}
\usepackage{amsmath}
\usepackage{amssymb}

\newcommand\Rey{\mbox{\textit{Re}}}  % Reynolds number

\newcommand\p{\partial}
\newcommand\commout[1]{}
\newcommand\bu{\mathbf u}
\newcommand\bU{\mathbf U}
\newcommand\bx{\mathbf x}
\newcommand\bv{\mathbf v}
\newcommand\bphi{\phi}
\newcommand\commentout[1]{}
\newcommand\pdd[2]{\frac{\p #1}{\p #2}}
\newcommand\mean[1]{\left\langle #1 \right\rangle}
\newcommand\DNS{direct numerical simulation}
\newcommand\bA{\mathbf A}
\newcommand\bB{\mathbf B}

\title{ Master-modes in 3D turbulent channel flow}

\author{S. I. Chernyshenko${}^1$ and M. E. Bondarenko${}^2$\\
${}^1$Department of Aeronautics, Imperial College London,\\
 Prince Consort Road,
London SW7 2AZ, UK;\\ 
${}^2$School of Engineering Sciences, University of Southampton,\\
Highfield,
Southampton,
SO17 1BJ,
UK }

\date{September 2008}

\begin{document}

\maketitle

\begin{abstract}
Turbulent flow fields can be expanded into a series in a set of basic functions. The terms of such series are often called modes. A master- (or determining) mode set is a subset of these modes,
 the time history of which uniquely determines the time history of the entire turbulent flow provided that this flow is developed. In the present work the existence of the master-mode-set is demonstrated numerically for turbulent channel flow. The minimal size of a master-mode set and the rate of the process of the recovery of the entire flow from the master-mode set history are estimated. The velocity field corresponding to the minimal master-mode set is found to be a good approximation for mean velocity in the entire flow field. Mean characteristics involving velocity derivatives deviate in a very close vicinity to the wall, while master-mode two-point correlations exhibit unrealistic oscillations. This can be improved by using a larger than minimal master-mode set. The near-wall streaks are found to be contained in the velocity field corresponding to the minimal master-mode set, and the same is true at least for the large-scale part of the longitudinal vorticity structure. A database containing the time history of a master-mode set is demonstrated to be an efficient tool for investigating rare events in turbulent flows. In particular, a travelling-wave-like object was identified on the basis of the analysis of the database.
Two master-mode-set databases of the time history of a turbulent channel flow are made available online at   http://www.dnsdata.afm.ses.soton.ac.uk/. The services provided include the facility for the code uploaded by a user to be run on the server with an access to the data.
\end{abstract}

\section{Introduction\protect\footnote{Experience of disseminating these results shows that for those who are new to the subject understanding can be achieved quicker if it is first  pointed out that master modes are not "something like POD". In fact, POD modes can also be divided into a master mode set and remaining  modes. It is also worth mentioning that master modes and determining modes are the same: we use the term master modes because we also need the complementary term slave modes.}}

Direct numerical simulations of turbulent flows are now quite common.
They have proved to be an important tool for fundamental research on turbulence, although calculating the entire turbulent flow around an aircraft is not yet possible. Indeed, even for moderate Reynolds numbers and simple geometries the computational effort required for \DNS\ is very high. 
For example, the recent simulation of a plane channel flow for $Re_{\tau}=2003$ ($Re_{\tau}$ is the Reynolds number based on friction velocity) required about $6\times 10^6$  processor-hours \cite[][]{HoyasJimenez2006}. Very few research groups have access to computational resources of that scale but even for them repeating a major \DNS\ is not always possible. It is natural, therefore, to store the data for future analysis, and indeed, databases of \DNS\ of turbulent flow are also becoming common. A turbulent flow database gives a considerable advantage, as the data can be quickly accessed at a moderate cost. However, a database of the full time history of a turbulent flow produced by a major \DNS\ would be very large. For example, the above-mentioned simulation gave about 25~TB of raw data.
Opening access (say, via Internet) to such a large volume is not feasible yet. For this reason the existing open-access \DNS\ databases contain only a limited number of samples of the instantaneous flow field complemented with various averaged data. Because of this the majority of researchers working on turbulence have a limited access to full direct numerical simulation results.

Direct numerical simulations are commonly performed for flows in bounded domains (say, a pipe flow periodic in axial direction). It is generally agreed that in this case as time advances the state of the flow approaches a finite-dimensional attractor.
\commentout{
\footnote{Rigorous results concerning this are available for 2D flow only, while in 3D \cite[see for example][]{ConstantinEtAl1985} a rigorous progress is slowed down by the absence of the proof of the existence of the solution of the Navier-Stokes equations. As far as numerical simulations are concerned this difficulty may be somewhat alleviated by the recent results on the existence of the solutions close to numerically calculated,  see \cite{ChernyshenkoEtAl2007} (currently available at arxiv.org as {\tt arXiv:math.NA/0607181}).}
}%commentout ends  
While the dimension of this attractor is believed to be high for high Reynolds numbers, it can be expected to be much smaller than the number of modes required for a fully resolved \DNS. Therefore, at least for developed turbulent flows it should be possible in principle to store only part of the modes used in the \DNS\ and still have a complete representation of the flow. This could resolve the problem of an efficient use of \DNS\ results by a wide research community. At the same time a study in this direction can give the answer to a deep fundamental question concerning the nature of turbulence, namely, the question about how much redundant information is indeed contained in a full time history of a turbulent flow. 

Determining the dimension of the attractor \cite[]{KeefeMoinKim1992} and storing the amplitudes of the number of modes slightly larger than this dimension does not actually resolve the issue of the size of a \DNS\ database. Indeed, since the attractor itself is unknown, even if the stored modes determine the state of the dynamical system on the attractor uniquely, there is no way to determine the amplitudes of the remaining modes. It is also not clear which modes should be stored. A more appropriate solution is to store a set of amplitudes of the so-called master modes, also known as determining modes. Let $\{\bphi_n(\bx)\}$ be a full basis so that a Navier-Stokes solution $\bu_A(t,\bx)$ can be expanded as 
$$
\bu_A=\sum_{n=1}^{\infty} \bA_n(t){\bphi_n(\bx)}.
$$
For double-periodic two-dimensional\footnote{For tree-dimensional flows obtaining similar results depends on the ability to prove that the vorticity is bounded, which is also needed for proving the existence of the tree-dimensional solution of the Navier-Stokes equations \cite[]{ConstantinEtAl1985}; this is a famous unsolved problem. However, for solutions obtained numerically a further step forward towards rigorous results can be made, see \cite[]{ChernyshenkoEtAl2007}.}  
                                                    flows \cite{FoiasProdi67} proved that there is a finite $N$ such that if another solution
$$
{\bu_B}=\sum_{n=1}^{\infty} \bB_n(t){\bphi_n(\bx)}
$$
has the same amplitudes of the first $N$ modes, that is if $\bA_n(t)=\bB_n(t)$ for $1\le n\le N$ then the two solutions converge with time, that is 

\begin{equation}\label{MasterModes}
||\bu_A-\bu_B||\to0\quad\mbox{as}\quad t\to\infty.
\end{equation}

 This implies that for a developed flow knowing only the time history of these $N$ amplitudes is sufficient for uniquely determining all the amplitudes. In other words, these $N$ master modes provide a full information about the developed turbulent flow.
The scope of the study described in the present paper includes clarifying further this idea as applied to numerical calculations, calculating the master-mode set for 3D channel flow, investigating some of the properties of this set, developing an online access to a database of the master-mode set, and giving an example of its use. 

\section{Master-mode set of a numerical solution}

\subsection{Definition}

In \DNS\ the solution is obtained by marching in time with discrete time steps. At each step the solution can be represented as
$$
\bu_i(\bx)=\sum_{n=1}^{S} \hat\bu_{i,n}{\bphi_n(\bx)},
$$
where the subscript $i$ corresponds to the solution at time $t_i,$ and $\bphi_n(\bx)$ is a given set of functions. Depending on the particular method, this set of functions can be a part of a full basis, as it is typical for Galerkin methods, and as is the case in the particular numerical method used in the present paper, or just an element of some interpolation formula, as in finite-difference methods based on a control volume approach. Each term in this sum is a mode, and $\hat\bu_{i,n}$ is the amplitude of the mode $n$ at the step $i.$ Then one time step can usually be described by the formula
\begin{equation}\label{TimeMarching}
\hat\bu_{i+1,n}=D_n(\hat\bu_{i,1},\hat \bu_{i,2},\dots, \hat \bu_{i,S}),\qquad n=1,\dots,S,
\end{equation}
where the functions $D_n$ depend on the particular numerical scheme. (Note that we consider the case when the boundary conditions and body forces are independent of time so that the functions $D_n$ are independent of $i.$ Note also that in some schemes $\bu_{i+1}$ depends not only on $\bu_i$ but also on $\bu_{i-1}.$ For simplicity we exclude such schemes from our analysis.) Let $\hat\bu_{i,n},i=1,2,\dots$ be a numerical solution satisfying~(\ref{TimeMarching}). For a given selection $n=m_1,m_2,...,m_K$ of the mode numbers consider a sequence $\{\bv_i(\bx)\}:$ 
$$
{\bv_i(\bx)}=\sum_{n=1}^{S} \hat\bv_{i,n}{\bphi_n(\bx)}
$$
such that
$$\hat\bv_{i+1,n}=\left\{
\begin{array}{lll}\hat\bu_{i+1,m_j}&  n=m_j,&j=1,\dots,K,\\
D_n(\bv_{i,1},\bv_{i,2}, \dots,\bv_{i,S}), & n\ne m_j,&j=1,\dots,K.\end{array}\right.
$$

In other words, values of $\hat\bv_{i+1,n}$ are obtained with the time-marching formula (\ref{TimeMarching}) applied to $\hat\bv_{i,n}$ if they do not belong to the set of modes in question and are taken to be equal to the corresponding modes of the numerical solution otherwise.

If for any $\bv_1(\bx)$
\begin{equation}\label{MMset}||\bv_i(\bx)- \bu_i(\bx)||\to0\quad\mbox{as}\quad i\to\infty\end{equation}
 then we will call the set $M$ of modes with $n=m_1,m_2,...,m_K$  the master-mode set of the numerical solution $\bu_i(\bx).$ It should be noted that the sequence $\bv_i(\bx)$ does not approximate a solution to the Navier-Stokes equations, in contrast to~(\ref{MasterModes}) where $\bu_A$ and $\bu_B$ are both solutions of the Navier-Stokes equations. It also should be noted that the selection $n=m_1,m_2,...,m_K$ giving a master-mode set for one solution does not necessarily give master-mode set for all other possible solutions of the same problem but with different initial conditions.

There is always a trivial master-mode set including all the modes of the numerical calculation. In this case $K=S.$ The natural goal, however, is to find master-mode sets with as small a $K$ as possible. It may well be that the minimal master-mode set is not unique: the property~(\ref{MMset}) is the property of the set of modes and not the property of individual modes. For this reason we use only the term master-mode set rather than the term master mode in the rest of the paper.

\subsection{Master code and slave code}

For any given set $M$ of modes there is an obvious way of verifying if this set is a master-mode set. The method consists in simultaneously running
two \DNS\ codes, which it is natural to call the master code and the slave code. The master code is just a standard \DNS\ code calculating the solution $\bu_{i,n}$. The slave code is an exact copy of the master code except that at each time step the amplitudes of the modes belonging to $M$ are replaced by their values calculated by the master code. If the initial conditions for both codes are identical, the entire master and slave solutions will be identical, too. If the initial conditions are different then the slave solution will tend to the master solution when set $M$ is a master-mode set. Otherwise, the solutions will deviate. 

If the modes are ordered by the likelihood of them belonging to the master-mode set then the minimal master-mode set can be found by dichotomy in the set size $K$.  
This approach was used by~\cite{OlsonTiti03}, \cite{HenshawKreissYstrom2003}, and \cite{YoshidaYamaguchiKaneda2005} in their studies of isotropic flows and proved to be highly efficient. In general, the result of such a calculation depends on the method of ordering the modes. In the studies on isotropic flows Fourier expansions were used and the modes were ordered by the magnitude of their wavenumbers. For a 3D channel flow a Fourier expansion in the wall-normal direction is not convenient, and the question of ordering the modes requires more attention, as described in the following section.

\section{Calculations of master-mode sets for 3D turbulent channel flow}

\subsection{Problem formulation}
Numerical calculations were performed for the flow in a plane channel. The governing equations of incompressible flow, that is the continuity equation and the momentum equations, were made non-dimensional with the channel half-width $h^{*}$ and the friction velocity $u_{\tau}^*=\sqrt{\tau_w^*/\rho^*}$, where  $\tau_w^*$ is the mean wall shear stress and $\rho^*$ is the density. Here asterisks are used to denote dimensional quantities. The Reynolds number is
$\Rey_\tau=u_{\tau}^* h^*/\nu^* $. The non-dimensional quantities are then defined as $ \bu=\bu^*/u_{\tau}^* $, $\bx=\bx^*/h^* $, $p=p^*/\rho^*
{u_\tau^*}^2,$ $t=t^*u_\tau^*/h^*.$ Using $u_{\tau}^*$ as the velocity scale results in the normalised mean pressure gradient $dP/dx=-1.$

The continuity equation has the form
$$
\nabla \cdot \bu=0,
$$
while the momentum equations are
\begin{equation}\label{NSE}
\frac{\partial \bu}{\partial t}+\bu\cdot\nabla\bu=-\nabla p - \textbf{e}_x
\frac{d P}{dx}
+\frac{1}{\Rey_\tau}\nabla^2\bu,
\end{equation}
where $\textbf{e}_x$ is the unit vector of the $x$-axis. In the above equation the pressure has been split into mean $P$ and fluctuating $p$ components. The coordinates are set as $x$ in the streamwise direction, $y$ in the spanwise direction and $z$ in the wall-normal direction, with the channel walls at $z=\pm1$. No-slip and impermeability boundary conditions are imposed on the walls. Periodicity conditions are used in the directions parallel to
the walls: $\bu(x+L_x,y,z,t)=\bu(x,y,z,t),$ $p(x+L_x,y,z,t)=p(x,y,z,t),$
$\bu(x,y+L_y,z,t)=\bu(x,y,z,t),$ and $p(x,y,z,t)=p(x,y+L_y,z,t),$ where $L_x$ and $L_y$ are the periods equal to the dimensions of the computation domain.

\subsection{The numerical method}

We used a version of the parallel pseudo-spectral channel flow code of \cite{SandhamHoward98} further modified for calculating master-modes.  The modification consisted in replacing parts of the code used in \cite{ChernyshenkoBaig05JFM} for calculating passive scalars with what effectively is another copy of the core Navier-Stokes solver, and organising the mechanism of replacing at each time step the amplitudes of a certain set of modes in the second solver by their amplitudes taken from the first solver. In order to ensure that no errors were introduced during the modification both solvers were then subject to all the standard tests for codes used for direct numerical simulation of turbulence.  

The \cite{SandhamHoward98} code uses Fourier series expansion in the wall-parallel directions and Chebyshev series in the wall-normal direction, so that the solution is represented in the form
\begin{equation}\label{Chebysh}
\bu{(x,t)}=\sum_{k_x=-N_x/2}^{N_x/2} \sum_{k_y=-N_y/2}^{N_y/2} \sum_{k_z=0}^{N_z-1} \hat{\bu}_{k_x,k_y,k_z}(t)e^{2\pi i(k_xx/L_x+k_yy/L_y)} T_{k_z}(z),
\end{equation}
where $ T_{k_z}(z)  $ is a Chebyshev polynomial of $k_z$ order and $\hat{\bu}_{k_x,k_y,k_z}(t)$ are complex-valued amplitudes of the modes. Since the solution is real-valued, $\hat{\bu}_{k_x,k_y,k_z}=\overline{\hat{\bu}}_{-k_x,-k_y,k_z}.$ Continuity imposes additional constraints on the components of the vectors $\hat{\bu}_{k_x,k_y,k_z}.$ In what follows, when the number of modes is given we will refer to the number of real-valued vectors: for example, with this way of counting modes the total number of modes in (\ref{Chebysh}) is $N_x\times N_y\times N_z.$ The table gives the parameters of the four main calculations referred to in the text, with some results described in more detail in what follows.

\begin{table}
\begin{center}
\begin{tabular}{|c|c|c|c|c|}\hline
$\Rey_\tau$ & $L_x \times L_y$ & $N_x\times N_y\times N_z$ & $\Delta t$ \\ 
\hline
85 &$5 \times 5$ &$32\times32\times 3$2& 0.0001\\
180& $4\times 3 $&$ 64\times64\times64 $& 0.001 \\
360 & $6\times 3$ & $128\times128\times160$ & 0.0005\\
360 & $14\times9$  & $512\times512\times 160$& 0.0005\\
\hline
\end{tabular}
\end{center}
\end{table}

\subsection{Mode ordering}\label{ModeOrdering}

Finding the smallest master-mode set by trial-and-error is inefficient. If modes are ordered, a more efficient dichotomy algorithm can be used, or simply a set of possible sizes of the master-mode set can be tried. Ordering the modes by the squared magnitude $k_x^2+k_y^2+k_z^2$ of their wavenumber is non-physical, since this method does not take into account the anisotropy of the flow and since $k_z,$ being the order of the Chebyshev polynomial, should not actually be directly compared with $k_x$ and $k_y.$ However, since this method is very straightforward it was tried. 

A more physical approach could be based on the averaged energy content of the modes. However, the modes (\ref{Chebysh}) are not orthogonal with respect to the standard energy norm. Instead, we introduce a Chebyshev-weighted energy norm:
$$||\bu||_C^2=\iiint \frac{|\bu|^2}{\sqrt{1-z^2}}\,dx\,dy\,dz.$$ This norm of the solution is equal to the sum of the norms of the modes, and hence the modes can be ordered by the time-averaged values of their amplitudes $\langle |\hat{\bu}_{k_x,k_y,k_z}|^2\rangle.$ The calculations were run for a certain period of time and statistics for calculating these averages was collected. Much longer calculation would be required for obtaining accurate values of the averages, but for our purpose high accuracy is not required. The modes were then ordered by $\langle |\hat{\bu}_{k_x,k_y,k_z}|^2\rangle$ with the assumption that more energetic modes are more likely to belong to the master-mode set.

\cite{Schmidtmann97,Schmidtmann98} in their work 
 on non-linear Galerkin methods suggested that
ordering the modes by their enstrophy content may be more efficient, and ordering the modes by their weighted enstrophy was also attempted in the present study.

Calculations performed for $\Rey_\tau =180$ in the box of the dimensions $L_x\times L_y\times L_z=4\times3\times2$ showed that the minimal size $N_{min}$ of the master-mode set  exceeds $5000$ if the modes are ordered by the wavenumber, that $N_{min}\sim4800$ for enstrophy-based ordering, and that $2800<N_{min}<3500$ for energy-based ordering. The energy-based ordering was then used in all other calculations.

\section{Properties of the master-mode set}

\subsection{Minimal size of a master-mode set}

The most widely-known estimate of the number of degrees of freedom of a turbulent three-dimensional flow follows from the arguments of \cite{LL59}:

\begin{equation}\label{LotoLd}N\sim(L_0/L_d)^3,\end{equation}
 where $L_0$ is the typical large lengthscale and $L_d$ is the Kolmogorov dissipation length given by
$$L_d=\nu^{3/4}/\varepsilon^{1/4}.$$
Here, $\nu$ is the kinematic viscosity and $\varepsilon$ is the energy dissipation rate per unit mass.  \cite{ConstantinEtAl1985} gave a rigorous mathematical derivation of~(\ref{LotoLd}) as an upper bound of the number of degrees of freedom. They also pointed out that in order to deduce the dependence of $N$ on the Reynolds number, an assumption is needed about the shape of the energy spectrum.  Assuming the Kolmogorov law of $5/3$ leads to the widely accepted estimate
\begin{equation}\label{Kolmogorov}N\sim\Rey^{9/4}.\end{equation}
However, in general the spectrum can have such a shape that the estimate becomes \begin{equation}\label{CubeLaw}N\sim\Rey^3.\end{equation} 
They then point out that since the Kolmogorov law is obtained for homogeneous isotropic high-Reynolds-number turbulence, (\ref{Kolmogorov}) can be invalid for relatively-low-$\Rey,$ inhomogeneous flows characteristic for direct numerical simulations of turbulence, and express a concern about the validity of numerical simulation for which the number of degrees of freedom is often predicted on the $\Rey^{9/4}$ rule. 

\begin{figure}
\centerline{\raisebox{0.25\textwidth}{$||\bu_m-\bu_s||$}\includegraphics[width=0.7\textwidth]{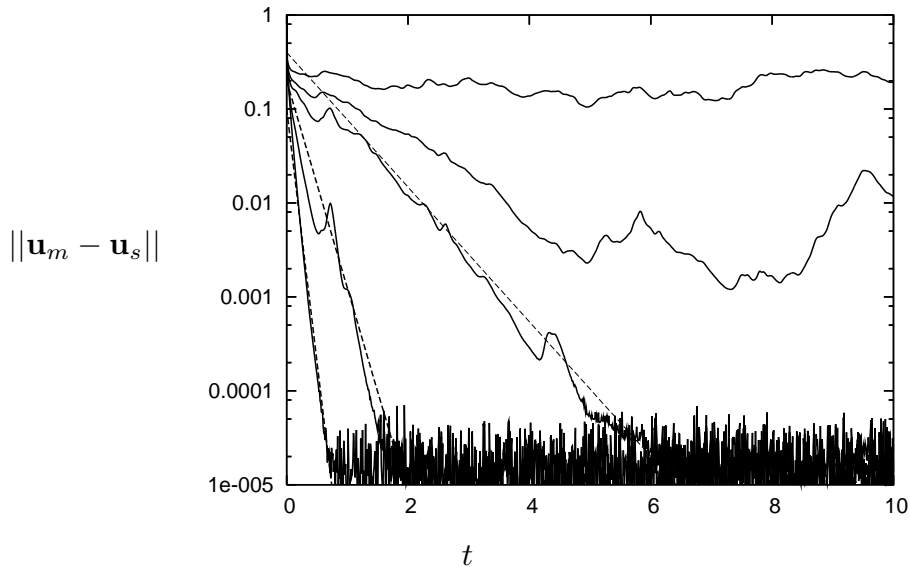}}
\centerline{$t$}
 \caption{Convergence history: $\Rey_\tau=180,$ $L_x=4,$ $L_y=3,$ $N_x\times N_y\times N_z=64\times64\times64,$ $N_{mm}=1600, 2800, 2900, 6000, 19000.$}\label{NormHistoryCaseOne} 
\end{figure}

On the other hand, \cite{HenshawKreissYstrom2003} argue that the number of degrees of freedom can be even less than (\ref{Kolmogorov}). They considered the case of decaying isotropic flow, which is difficult to compare directly to the developed channel flow considered in the present study. Their main idea (expressed using our terminology) is that a master-mode flow field can form regions of high dissipation (shear layers or concentrated vortices) and it can advect the errors in the slave-mode field to these high-dissipation regions where the errors will quickly decay. Both the advection rate and the decay rate may be even independent of $\Rey$ (the latter because as $\Rey$ increases the thickness of, say, shear layers can decrease accordingly). This was demonstrated to be sometimes the case for Burgers' equations. For the Navier-Stokes equations, however, \cite{HenshawKreissYstrom2003} observed an increase in the number of degrees of freedom as viscosity decreased, but the rate of this increase was not evaluated.
 
\cite{YoshidaYamaguchiKaneda2005} considered master modes in a forced isotropic flows and their calculations confirm (\ref{Kolmogorov}). However, this does not refute the \cite{ConstantinEtAl1985} argument, since one of the possible reasons for the deviation from (\ref{Kolmogorov}) is the flow anisotropy.    

\begin{figure}
\centerline{\raisebox{0.25\textwidth}{$||\bu_m-\bu_s||$}\includegraphics[width=0.7\textwidth]{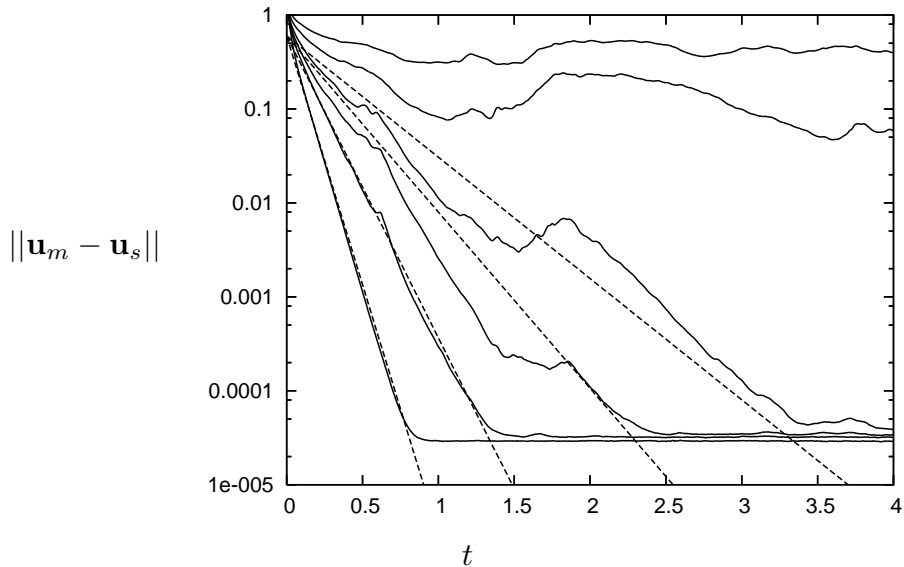}}
\centerline{$t$}
 \caption{Convergence history: $\Rey_\tau=360,$ $L_x=6,$ $L_y=3,$ $N_x\times N_y\times N_z=128\times128\times160,$ $ N_{mm}=18365, 22879, 29716, 34601, 41069, 63448. $}\label{NormHistoryCaseThree} 
\end{figure}

Calculating the minimum size of the master-mode set is an efficient way of estimating the number of the degrees of freedom in the turbulent flow, thus addressing the question of which rule, (\ref{Kolmogorov}) or (\ref{CubeLaw}), is the more appropriate for the channel flow. Figures~\ref{NormHistoryCaseOne} and~\ref{NormHistoryCaseThree} show the time-dependence of the $L_2$ norm of the difference between the master-code solution $\bu_m$ and the slave-code solution $\bu_s$ for different sizes $N_{mm}$ of the master-mode set for two of the calculated cases. The calculations were performed with single precision, so that the random oscillations at the bottom of figure~\ref{NormHistoryCaseOne} are the result of the rounding error noise. In figure~\ref{NormHistoryCaseThree} the effect of rounding errors is averaged over a larger volume and oscillates less. These figures give the bounds for the minimum size of the master-mode set, namely, for $\Rey_\tau=180$ case  $2800<N_{min}<2900$ (figure~\ref{NormHistoryCaseOne}), and for $\Rey_\tau=360$ case  $22879<N_{min}<29716$ (figure~\ref{NormHistoryCaseThree}).

Note that the convergence is not monotone for smaller $N_{mm}$ and, hence, the convergence rate (determined from the slope of the dashed lines in the following analysis) can be found only approximately. 

Figures~\ref{NormHistoryCaseOne} and \ref{NormHistoryCaseThree} establish the existence of the master-mode sets in three-dimensional turbulent flow in a plane channel.

We also checked that increasing the number of modes used in the numerical calculations above the necessary minimum does not affect the size of the
minimal master-mode set. Namely, calculations for $\Rey_\tau=360,$ $L_x=6,$ $L_y=3$ were performed with an increased number of modes $256 \times 256 \times 160.$ The minimal size of the master-mode set was found to be about 29000, which is the same as the minimal size of the master-mode set for $128\times128\times160$ modes calculation.

Two other calculations were also performed, for $\Rey_\tau=85,$ $L_x=5,$ $L_y=5,$ $N_x\times N_y\times N_z=32\times32\times32,$ which gave $595<N_{min}<937,$ and for $\Rey_\tau=360,$ $L_x=14,$ $L_y=9,$ $N_x\times N_y\times N_z=512\times512\times160,$ which gave $161336<N_{min}<225900.$ 
The calculations with $\Rey_\tau=85$ were performed for the sake of comparisons with the only other known previous attempt to calculate the number of the degrees of freedom of a plane channel flow. \cite{KeefeMoinKim1992} obtained an approximate estimate $D_\lambda\approx780$ of the Lyapunov dimension $D_\lambda$ of the attractor for a plain channel flow in a box of the size $L_x=L_y=1.6\pi$ using two calculations with two grids, $16\times33\times8$ and $32\times32\times32.$ For the latter grid only partial results were obtained.  We had to use a slightly higher Reynolds number since at $\Rey_\tau=80$ we were unable to obtain a turbulent flow. Note that the number of degrees of freedom corresponds to twice the size of the minimal master-mode set since in our notation each mode represents two degrees of freedom, the mode amplitude being a vector the three components of which are related by the continuity equation.

The Lyapunov dimension of the attractor and the minimal size of the master-mode set are not the same but they both provide a measure of the number of the degrees of freedom of the flow. In view of this and all other factors the agreement between \cite{KeefeMoinKim1992} and our results is quite good. The case for $\Rey_\tau=360,$ $L_x=14,$ $L_y=9$ was selected as the largest calculation for which we could produce a master-mode set time-history database with the available computational resources.

\begin{figure}
\centerline{\raisebox{0.25\textwidth}{$\displaystyle \frac{N_{min}}{L_xL_y}$}\includegraphics[width=0.7\textwidth]{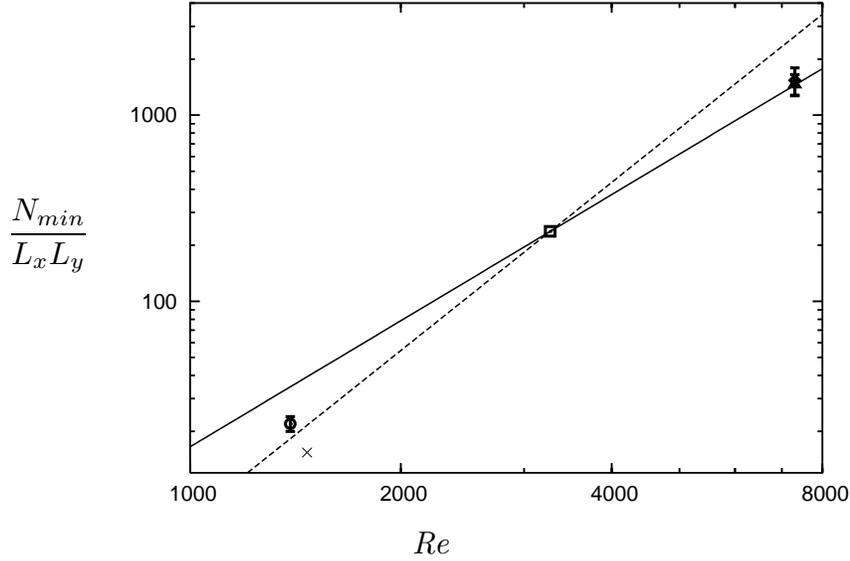}}
\centerline{$\Rey$}
 \caption{Error bounds on the size of the minimum master-mode set for calculated cases and two estimates, $\Rey^{9/4}$ (solid line) and $\Rey^3$ (dashed line). $\Rey_\tau=85,$ $L_x\times L_y=5\times5,$ $\circ;$  $\Rey_\tau=180,$ $L_x\times L_y=4\times 3,$  ${\Box}$ (error bounds too close to be seen);  $\Rey_\tau=360,$ $L_x\times L_y=6\times 3,$ ${\triangle};$  $\Rey_\tau=360,$ $L_x\times L_y=14\times9,$ ${ \diamond}$ (overlaps with the previous case); \cite{KeefeMoinKim1992}, $\times.$}
\label{Dimension} 
\end{figure}

Obviously, the number of the degrees of freedom of the flow is dependent on the dimensions of the computational box. The easiest and most natural assumption is that a factor $L_xL_y$ should be included in the right-hand-side of (\ref{Kolmogorov}) and (\ref{CubeLaw}). Note that $L_x$ and $L_y$ are the non-dimensional ratios of the length and depth of the computational box to the half-width of the channel.  Also, $\Rey$ in these formulae should be based on the maximum velocity (see (28) in \cite{ConstantinEtAl1985}). Hence, figure~\ref{Dimension} shows a plot of $N_{min}/(L_xL_y)$ versus $\Rey$ based on the maximum velocity for all available data together with lines corresponding to (\ref{Kolmogorov}) and (\ref{CubeLaw}). (The \cite{KeefeMoinKim1992} $\Rey_\tau=80$ point is on the right from our $\Rey_\tau=85$ point because the maximum mean velocity in their calculation was somewhat greater than in our calculations.) The available data fit (\ref{CubeLaw}) for smaller $\Rey$ and (\ref{Kolmogorov}) for larger $\Rey.$ This suggests that the \cite{ConstantinEtAl1985}) argument may be correct for smaller $\Rey,$ for which it was proposed, while the Kolmogorov spectrum is emerging at larger $\Rey.$ More data is needed for full certainty.

\subsection{Convergence rate}

The rate at which the difference between the master solution and the slave solution tends to zero can be estimated in the following way. Let $N_{mm}$ be the number of the modes whose amplitudes are being passed from the master to the slave code. When $N_{mm}>N_{min}$, the difference between the master and the slave solution converges exponentially, $||\bu_m-\bu_s||\sim e^{-\alpha t}$ \cite[]{OlsonTiti03}. The value of $\alpha$ depends on $N_{mm}$. Assuming that the decay is due to viscous effects gives an estimate $\alpha\sim
L_m^{-2}\Rey_{\tau}^{-1},$ where $L_m$ is the largest characteristic length scale of the slave mode set (non-dimensional quantities corresponding to (\ref{NSE}) are used throughout). One can assume that $L_m$ is proportional to the cubic root of the volume of the computational box divided by the number of modes: $L_m\approx \mathrm{const} (L_xL_yL_z/N_{mm})^{1/3}.$ This gives the formula
\begin{equation}\label{alpha} \alpha=\frac{C}{\Rey_\tau }\left(\frac{ N_{mm}}{L_xL_yL_z}\right)^{2/3}.\end{equation}
It is equivalent to $\alpha^*{L_m^*}^2/\nu=\alpha^*({L_x}^*{L_y}^*L_z^*/N_{mm})^{2/3}/\nu =\mathrm{const,}$ where ${}^*$ denotes dimensional quantities and $\nu$ is the kinematic viscosity. Naturally, as $N_{mm}$ approaches $N_{min},$ $\alpha$ should tend to zero. This is not reflected by~(\ref{alpha}), which, therefore, can be expected to describe an asymptote approached from below as $N_{mm}\to\infty.$

\begin{figure}
\centerline{\raisebox{0.25\textwidth}{$\alpha\Rey_\tau$}\includegraphics[width=0.7\textwidth]{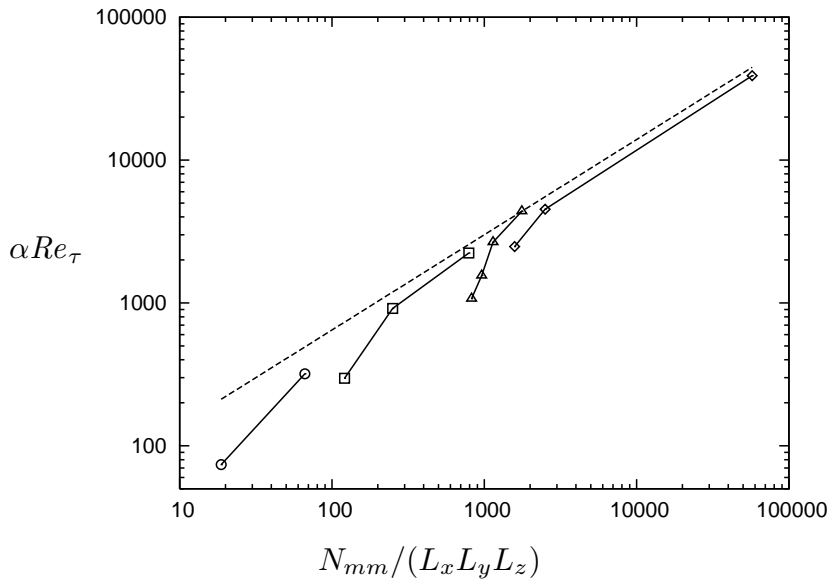}}
\centerline{$N_{mm}/(L_xL_yL_z)$}
 \caption{Convergence rate, $L_z=2:$ $\Rey_\tau=85,$ $L_x\times L_y=5\times5,$ $\circ;$  $\Rey_\tau=180,$ $L_x\times L_y=4\times 3,$  ${\Box};$  $\Rey_\tau=360,$ $L_x\times L_y=6\times 3,$ ${\triangle};$  $\Rey_\tau=360,$ $L_x\times L_y=14\times9,$ ${ \diamond}.$ }\label{Convergence}
\end{figure}

Figure~\ref{Convergence} shows the values of $\alpha\Rey_\tau$ plotted against $N_{mm}/(L_xL_yL_z)$ for all calculated cases together with the line given by~(\ref{alpha}) with $C=30.$ The values of $\alpha$ were obtained by estimating the slope of the plot of the $L_2$ norm (that is the square root of the integral of $|\bu_m-\bu_s|^2$ over the calculation domain) versus time. It appears that (\ref{alpha}) can indeed be used for rough estimates of the convergence rate at large $N_{mm}$.

\cite{YoshidaYamaguchiKaneda2005} found that the convergence rate multiplied by the Kolmogorov time scale and plotted versus the cut-off wavenumber multiplied by the Kolmogorov length is independent of the Reynolds number. In a channel flow the energy dissipation rate depends on the distance to the wall. We can, however, introduce an average energy dissipation rate (per unit volume) as $\epsilon=U_{mean},$ where $U_{mean}$ is the mean velocity  (note that due to the scaling used in our work the pressure gradient is equal to $-1$, and all the quantities are non-dimensional). We then introduce the Kolmogorov time and length scales as $\tau=(\epsilon\Rey_\tau)^{-1/2}$ and $\eta=(\epsilon\Rey_{\tau}^3)^{-1/4}.$ In \cite{YoshidaYamaguchiKaneda2005} the number of modes per unit volume corresponding to the cut-off wave number $k_a$ can be evaluated as $(4/3)\pi k_a^3/(2\pi)^3$ while in our case it is $N_{mm}/(L_xL_yL_z).$ Accordingly, we define our cut-off wavenumber as

$$k_a=2\pi\left(\frac{3N_{mm}}{4\pi L_xL_yL_z}\right)^{1/3}.$$

\begin{figure}
\centerline{\raisebox{0.25\textwidth}{$\tilde\alpha $}\includegraphics[width=0.7\textwidth]{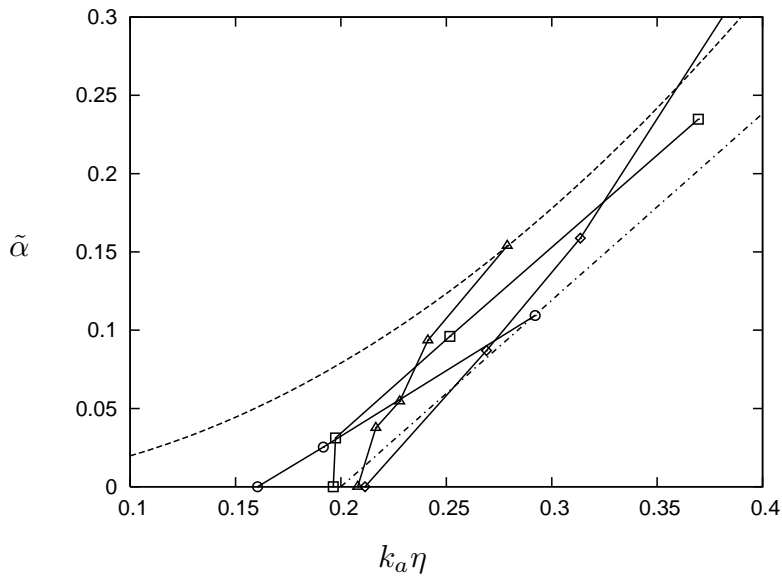}}
\centerline{$k_a\eta$}
 \caption{Convergence rate. The dash-dotted line is an approximation of the \cite{YoshidaYamaguchiKaneda2005} results. Other lines and symbols are the same as in figure~\ref{Convergence}. \label{ConvergenceTwo}}
\end{figure}

Figure~\ref{ConvergenceTwo} shows the dependence of $\tilde\alpha=\alpha\tau$ on $k_a\eta.$ The dash-dotted line is an approximation of the \cite{YoshidaYamaguchiKaneda2005} results. Of course this comparison is only tentative since the channel flow is not isotropic and homogeneous. However, our results at least do not contradict the conclusion made in \cite{YoshidaYamaguchiKaneda2005} that the Kolmogorov-scaled convergence rate is determined by the quantities for the energy dissipation scale rather than the energy containing scale.

In large eddy simulation approach the full solution is also divided into large scale and small scale components. Then the large scale motion is simulated numerically with the effect of small scale on large scale described with a subgrid-scale model. If a similarity between the large scale motion in the large eddy simulation approach and master-modes is assumed then from the results for master modes it follows that subgrid scale models should have memory, for example in the form of relaxation relationships. The above data on the convergence rate can be used to estimate the time constants of these relaxation relationships.

\subsection{Mean flow}

\begin{figure}
\centerline{\raisebox{0.25\textwidth}{$\langle u \rangle$ }\includegraphics[width=0.7\textwidth]{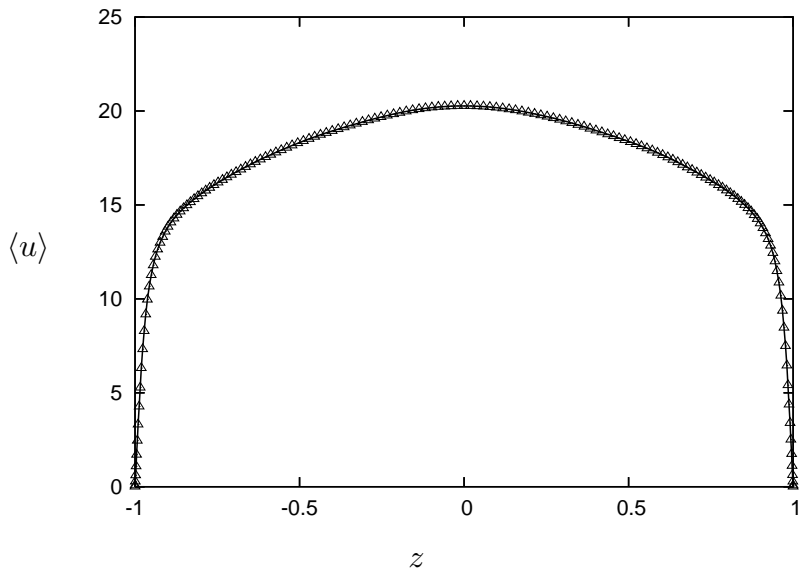}}
\centerline{$z$}
 \caption{Mean velocity, $\Rey_\tau=360,$ $L_x\times L_y=6\times 3,$ full flow (solid curve) and a minimal master-mode set ($N_{mm}=29716$), $\triangle.$}\label{Mean}
\end{figure}

A master-mode set contains full information about the particular realisation of the developed turbulent flow, since there is a method for recovery of the rest of the modes of that realisation. This makes the master-mode set a natural choice for storing the time history of the flow in a database. Of course, the velocity field corresponding to the master-mode set alone does not exactly coincide with the flow. However, for applications where the approximate representation given by the master-mode set itself is sufficient, using it is convenient and gives a considerable saving of computing resources. Figure~\ref{Mean} shows the comparison of the mean velocity profile obtained from the master-mode set alone and the velocity profile of the full (that is obtained from all the modes used in the direct numerical simulation) flow for $\Rey=360,$ $L_x=6,$ $L_y=3$ and the master-mode set size $N_{mm}=29716,$ which is close to the minimal master-mode set size. 

\subsection{Turbulent kinetic energy budget}

For a channel flow

$$\pdd{\langle \bu'^2\rangle}t=P+T+D+\Pi+\Phi,$$
where
$\displaystyle P=-2\mean{u'w'}\pdd{\mean{u}}z,$
$\displaystyle T=-\pdd{\mean{w'u'_i u'_i}}z,$
$\displaystyle D=\frac1{\Rey_{\tau}}\pdd{^2\mean{ u'_i u'_i}}{z^2},$
$\displaystyle \Pi=-2\mean{u'_i\pdd{p'}{x_i}},$ and
$\displaystyle \Phi=-\frac2{\Rey_\tau}\mean{\pdd{u'_i}{x_j}\pdd{u'_i}{x_j}}.$
The angular brackets denote averaging. Double notation is used with $x=x_1,$ $y=x_2,$ $z=x_3,$ $u=u_1,$ $v=u_2,$ and $w=u_3,$ with summation over repeated indices.

\begin{figure}
\centerline{\raisebox{0.25\textwidth}{$P,T,D,\Phi$}
\includegraphics[width=0.7\textwidth]{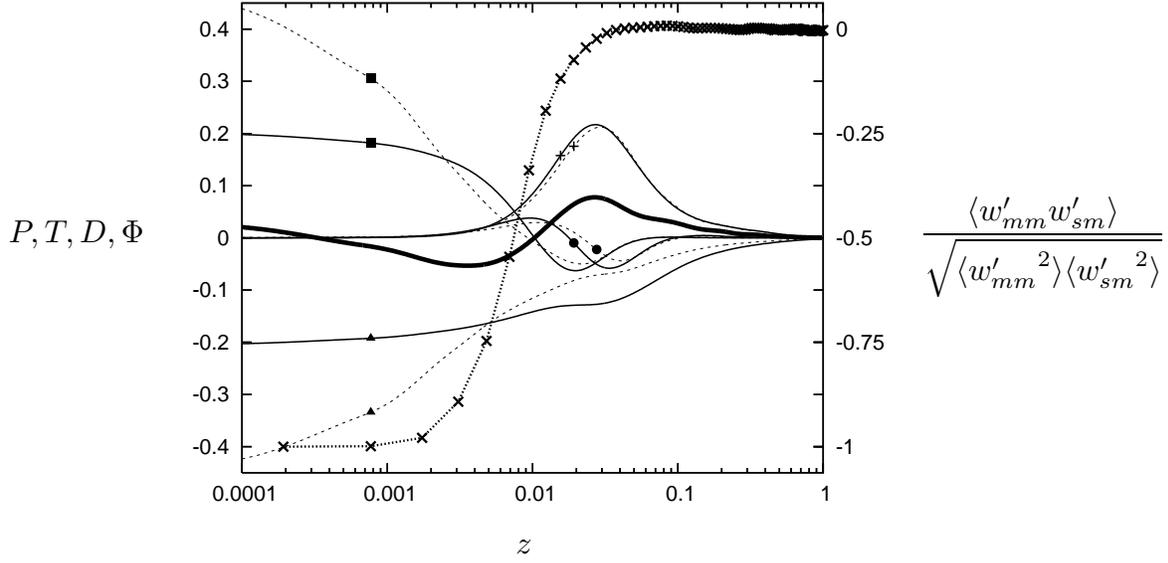}\raisebox{0.25\textwidth}{$\displaystyle\frac{\langle w'_{mm}w'_{sm} \rangle}{\sqrt{\langle {w'_{mm}}^2\rangle\langle {w'_{sm}}^2\rangle}}$}
}
\centerline{$z$\qquad\qquad\ }
\caption{Turbulent kinetic energy budget components: full flow (solid curves) and master-mode set only (dashed curves); production $P$, +; transport $T$, $\bullet;$ viscous diffusion $D$, $\blacksquare;$ viscous dissipation, $E$ $\blacktriangle;$ total budget for the master-mode set (thick solid line). Correlation between the master-mode set and slave-mode set wall-normal velocity fields (thick dashed curve with $\times$).  $\Rey_\tau=360,$ $L_x\times L_y=6\times 3,$ $N_{mm}=29716.$ \label{Budgets}} 
\end{figure}

Figure~\ref{Budgets} shows the various components of the turbulent energy budget for the same calculation and the same master-mode set. The pressure term $\Pi$ is not shown since it is small, however for $\Pi$ the agreement between the minimal master-mode set field and the full flow is very good.
Note the use of the logarithmic scale. There is a significant deviation of the master-mode set curves from the full flow results. This deviation, however, is limited to a few near-wall points. The reason for this discrepancy becomes evident from the plot of the correlation between the minimal master-mode set velocity $\bu_{mm}$ and the slave-mode set velocity $\bu_{sm}$ (not to be confused with $\bu_m$ and $\bu_s$), also given in the same figure for the wall-normal velocity component. Away from the wall the correlation is small, but near the wall the correlation coefficient is close to $-1.$ Regression analysis shows that close to the wall $\bu_{sm}\approx - \bu_{mm},$ while the full velocity $\bu=\bu_{mm}+\bu_{sm}$ is significantly smaller in magnitude than both $\bu_{mm}$ and $\bu_{sm}.$ This happens because the Chebyshev polynomials, which consitute the modes,  strongly oscillate near the wall, while the full flow velocity does not. This is another indication of the well-known fact that Chebyshev polynomials over-resolve turbulent flow near the wall if they fully resolve it away from the wall. One can obtain the same conclusion using, for example, the so-called directional dissipation scale \cite[]{Manhart2000}. The amplitude of the modes containing lower-order Chebyshev polynomials is determined by the solution behaviour over the channel-width scale. The near-wall over-resolving property results in the level of derivatives corresponding to these modes being well above the magnitude of the derivatives of the solution. Near the wall the higher-order modes, therefore, have to cancel out the lower-order modes. This phenomenon manifests itself when the solution is divided into a master mode set and a slave mode set.  Probably, using the high correlation between master and slave modes near the wall the agreement can be improved; away from the wall the minimal master-mode set provides a reasonable approximation for energy budget components. 

\subsection{Two-point correlations}

For any given wall-parallel plane with coordinate $z$ in a plane channel two-point correlation coefficients can be defined as
$$R_{ij}(\Delta x, \Delta y,z) = \frac{\mean{u'_i(x,y,z) u'_j(x+\Delta x,y+\Delta y,z)}}{
\sqrt{\mean{{u'_i}^2(x,y,z)}\mean{{u'_j}^2(x,y,z)}}}.$$
This definition exploits the homogeneity in the wall-parallel $x$ and $y$ directions.  

\begin{figure}
\centerline{\raisebox{0.25\textwidth}{$R_{ii}$}
\includegraphics[width=0.7\textwidth]{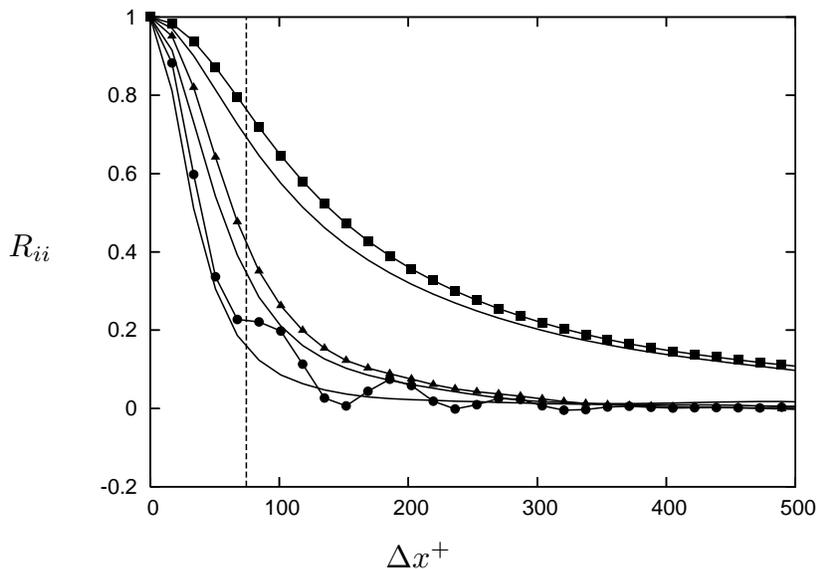}}
\centerline{$\Delta x^+$}
\caption{Streamwise ($\Delta y=0$) two-point correlations, curves without symbols, full flow; curves with symbols, master-mode-set; longitudinal velocity correlation, $R_{uu},$ $\blacksquare;$ spanwise velocity correlation, $R_{vv},$ $\blacktriangle;$ wall-normal velocity correlation, $R_{ww}$, $\bullet.$ For $\Rey_\tau=360,$ $z^+=5.6,$ $L_x\times L_y=6\times3,$ and minimal master-mode set ($N_m=2900.$) \label{StreamwiseCor}} 
\end{figure}

\begin{figure}
\centerline{\raisebox{0.25\textwidth}{$E$}
\includegraphics[width=0.7\textwidth]{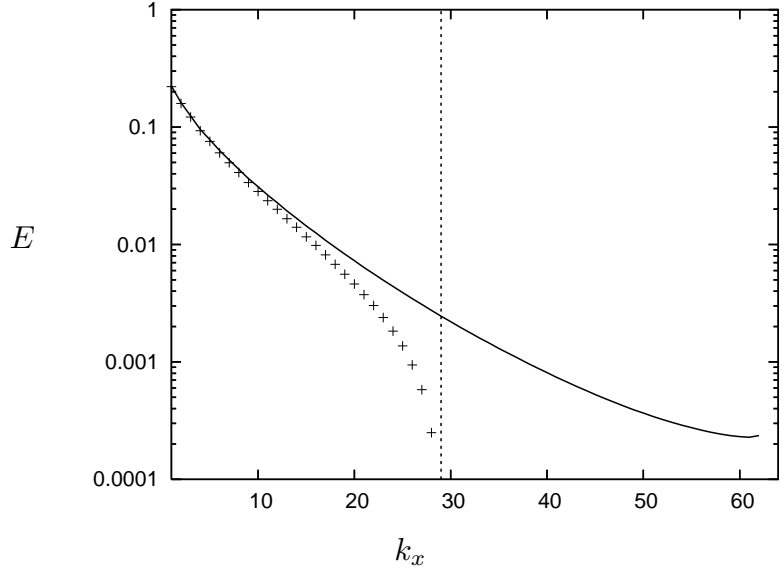}}
\centerline{$k_x$}
\caption{Streamwise energy spectrum $E=\mean{\left|\sum_{k_y=-N_y/2}^{N_y/2} \sum_{k_z=0}^{N_z-1} \hat{\bu}_{k_x,k_y,k_z}(t)e^{2\pi ik_yy/L_y} T_{k_z}(z)\right|^2}$, see~(\ref{Chebysh}). Full flow, the curve; master-mode-set, +. For $\Rey_\tau=360,$ $z^+=5.6,$ $L_x\times L_y=6\times3,$ and the minimal master-mode set ($N_{mm}=29716$). \label{Spectrum}} 
\end{figure}

Figure~\ref{StreamwiseCor} shows the streamwise ($\Delta y=0$) two-point correlation coefficients for the full flow and the minimal master-mode set velocity field. The full-flow and the master-mode-set curves for the longitudinal velocity and the spanwise velocity correlations are in a reasonable agreement. However, the correlation for the master-mode-set wall-normal velocity exhibits oscillations around the respective full-flow correlation for $\Delta x$ greater than about 75. This can be caused by the sharp cut-off in the energy spectrum of the master-mode-set field at approximately the same length scale\footnote{This explanation was suggested by Prof. I.P.Castro}. 
The cut-off scale is indicated by a vertical dashed line in figure~\ref{StreamwiseCor} and in figure~\ref{Spectrum}, which shows the corresponding energy spectrum. The relation between the length scale and the wave number, $x^+=360L_x/k_x,$ follows from~(\ref{Chebysh}). 

\begin{figure}
\centerline{\raisebox{0.25\textwidth}{$R_{ii}$}
\includegraphics[width=0.7\textwidth]{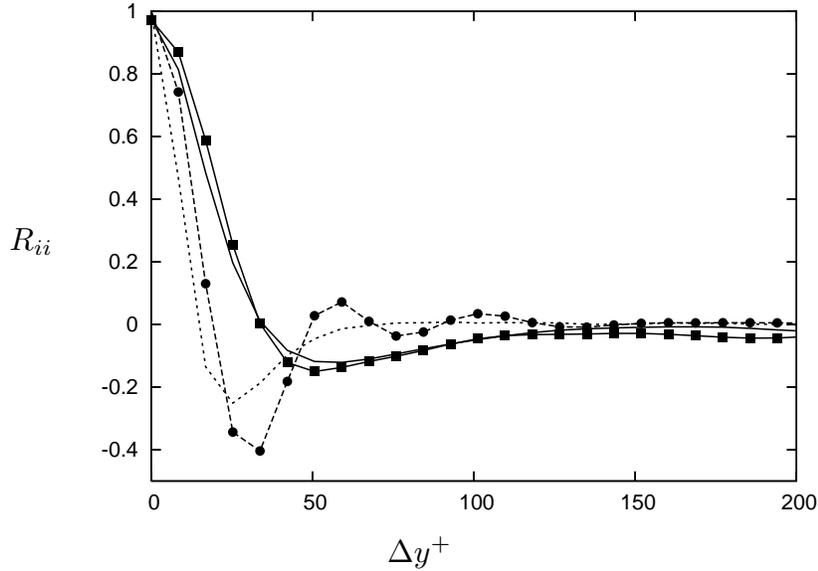}}
\centerline{$\Delta y^+$}
\caption{Spanwise ($\Delta x=0$) two-point correlations, curves without symbols, full flow; curves with symbols, master-mode-set; longitudinal velocity correlation, $R_{uu},$ $\blacksquare;$  wall-normal velocity correlation, $R_{ww}$, $\bullet.$ For $\Rey_\tau=360,$ $z^+=5.6,$ $L_x\times L_y=6\times3,$ and minimal master-mode set ($N_{mm}=29716$). \label{SpanwiseCor}} 
\end{figure}

For the spanwise correlation (figure~\ref{SpanwiseCor}) the master-mode-set correlation begins to oscillate around the full-flow correlation at smaller scale, which may indicate that the cut-off level in the spanwise spectrum is at this smaller scale. The correlation for the spanwise velocity component is not shown because it is very close to the correlation for the longitudinal velocity component. Note that the oscillations are observed for the wall-normal component only. This may be due to the fact that at this distance to the wall ($z^+=5.6$) the wall-normal velocity component is much smaller than the other velocity components: the slave modes, being selected by an energy criterion (see section~\ref{ModeOrdering}) may give a relatively higher contribution to the wall-normal velocity component. Naturally, increasing the number of modes in the master-mode set reduces the oscillations, see section~\ref{MMdatabase}. 

\subsection{Structures}

While the master-mode set contains full information about the developed turbulent flow in the sense that its time history uniquely identifies the particular realisation of the turbulent flow, it is far more difficult to establish whether the master-mode-set velocity field contains all the essential flow dynamics. If this is believed to be true then considering which organised structures are observed in the minimal master-mode-set and which are observed in the remaining modes may help to establish a relative importance of different organised structures. On the other hand, if the relative importance of different structures is considered to be known, the analysis of the structures in the minimal master-mode-set and the remaining modes can be used to estimate the dynamical significance of the minimal master-mode set. From the results obtained so far it appears, however, that the minimal master-mode set contains almost all the structures.

\begin{figure}
\centerline{
\includegraphics[width=0.33\textwidth]{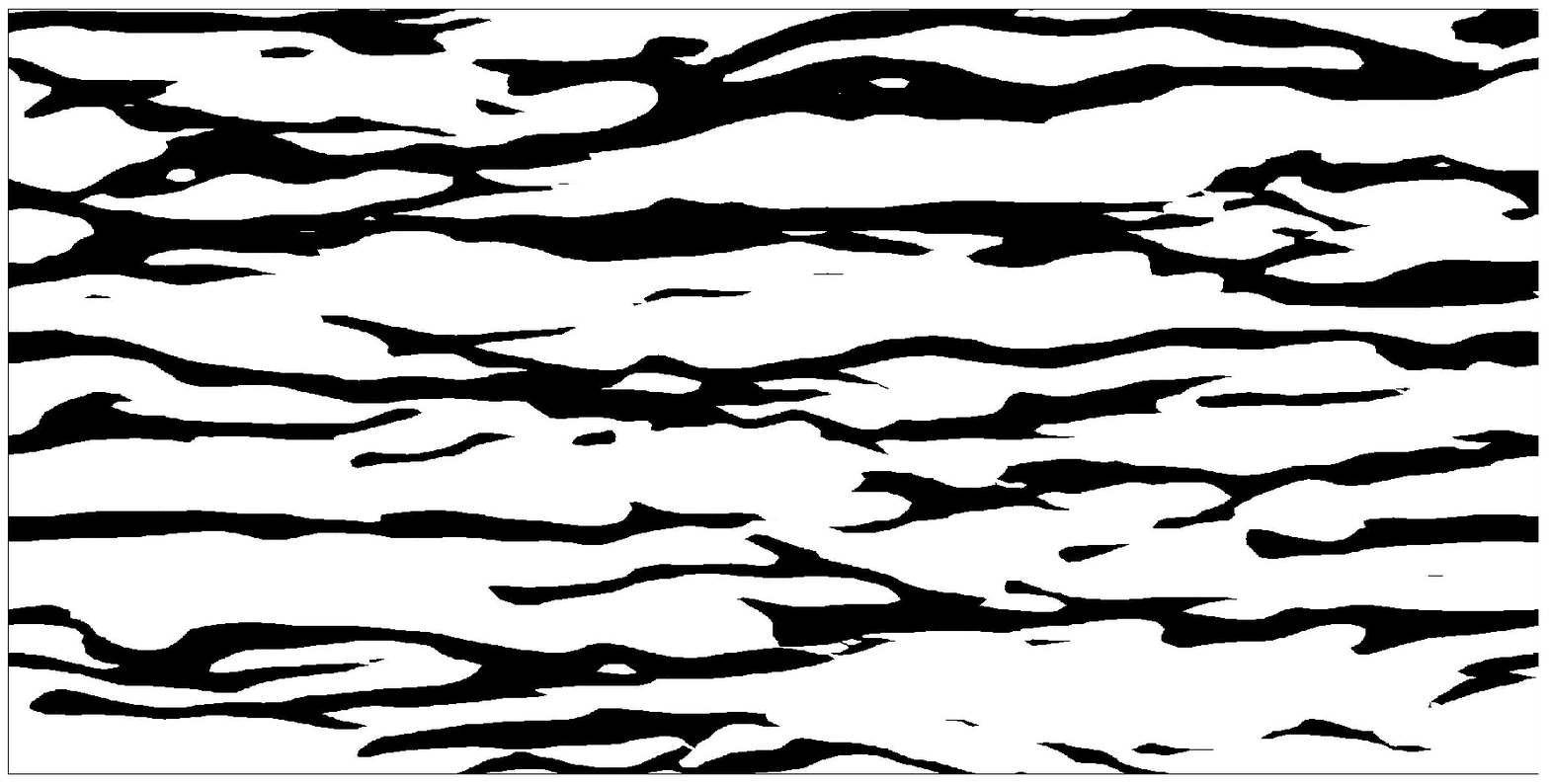} 
\includegraphics[width=0.33\textwidth]{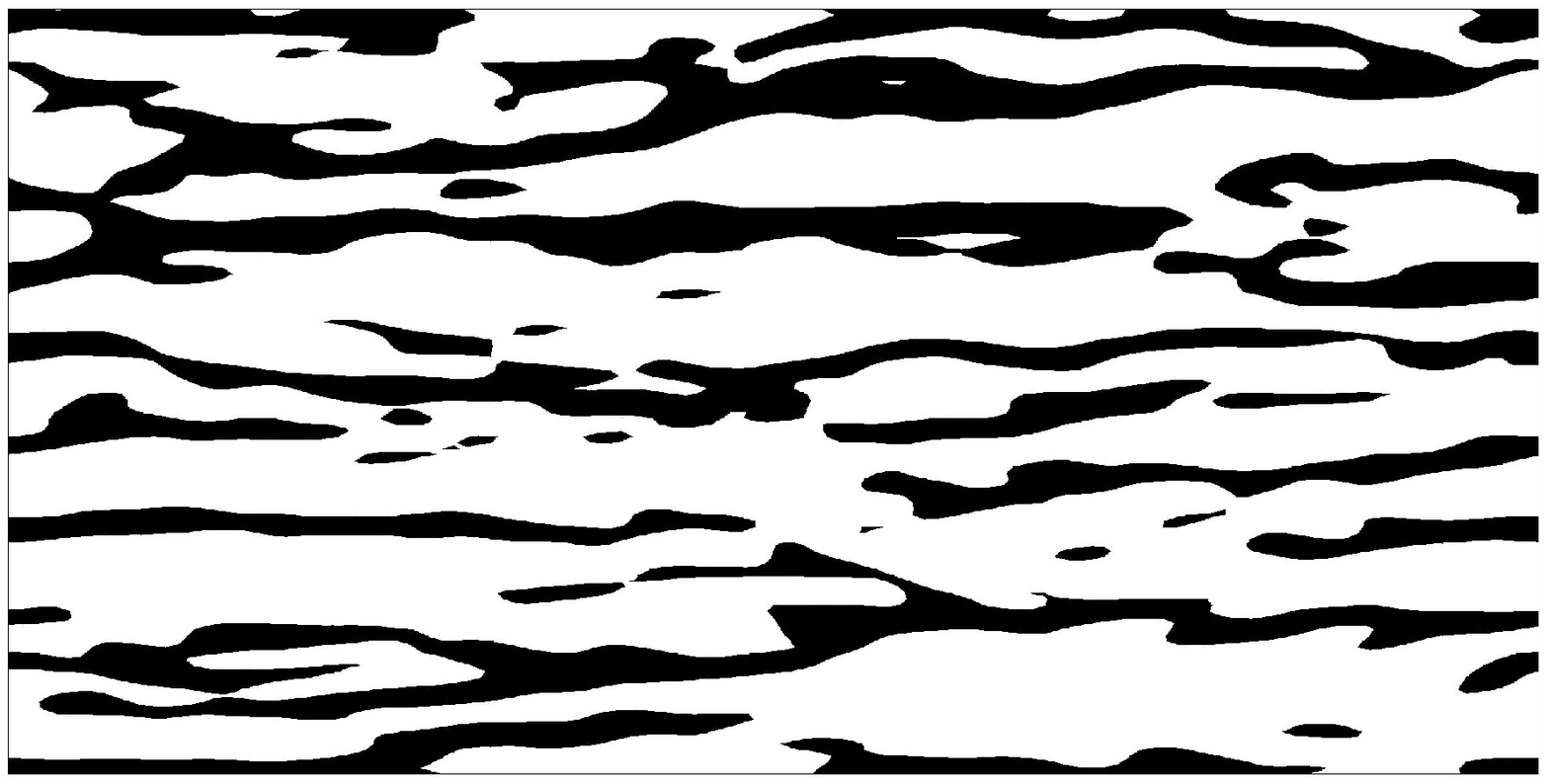}
\includegraphics[width=0.33\textwidth]{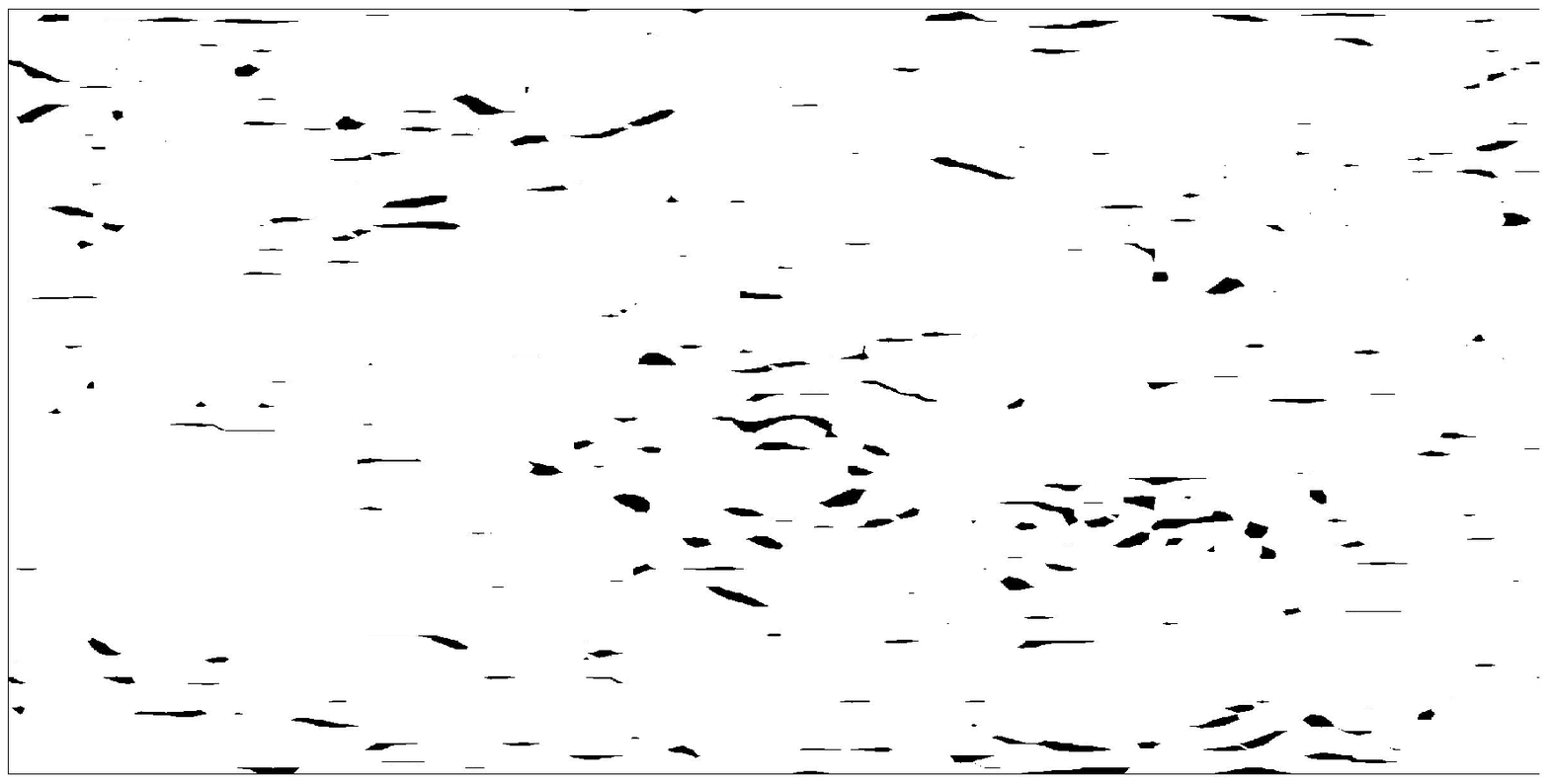}}
\centerline{\it a)\hspace{0.3\textwidth} b) \hspace{0.3\textwidth} c)}

\ 

\centerline{
\includegraphics[width=0.33\textwidth]{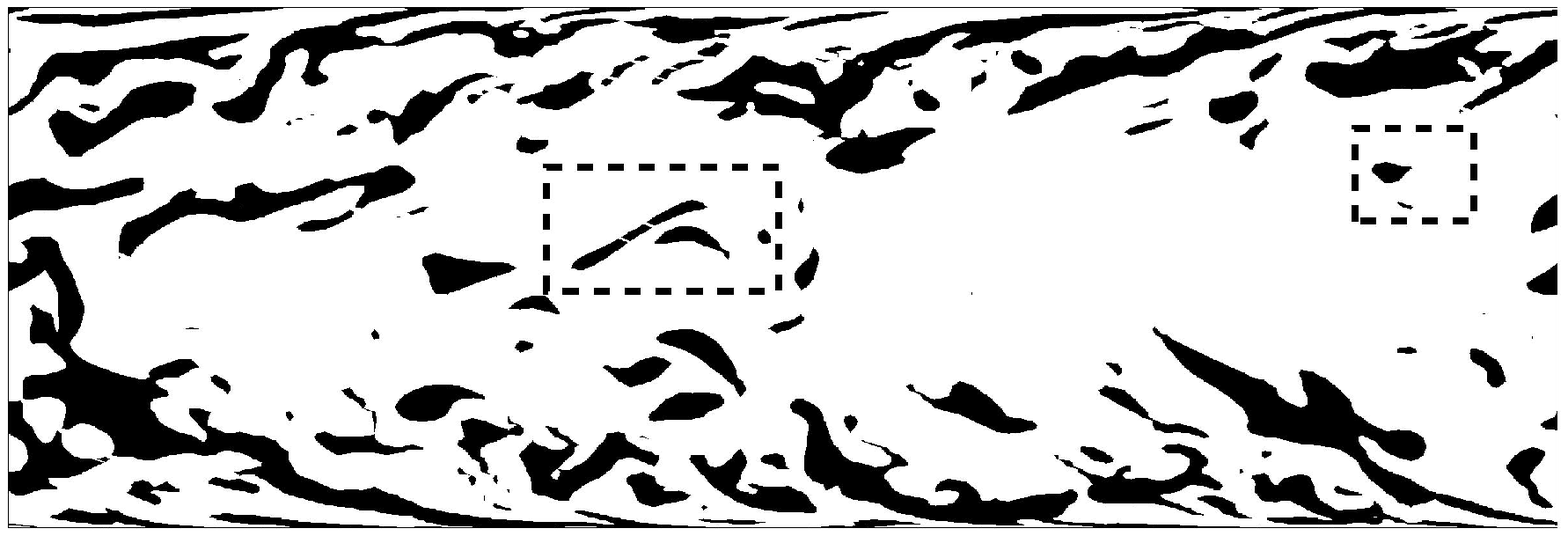}
\includegraphics[width=0.33\textwidth]{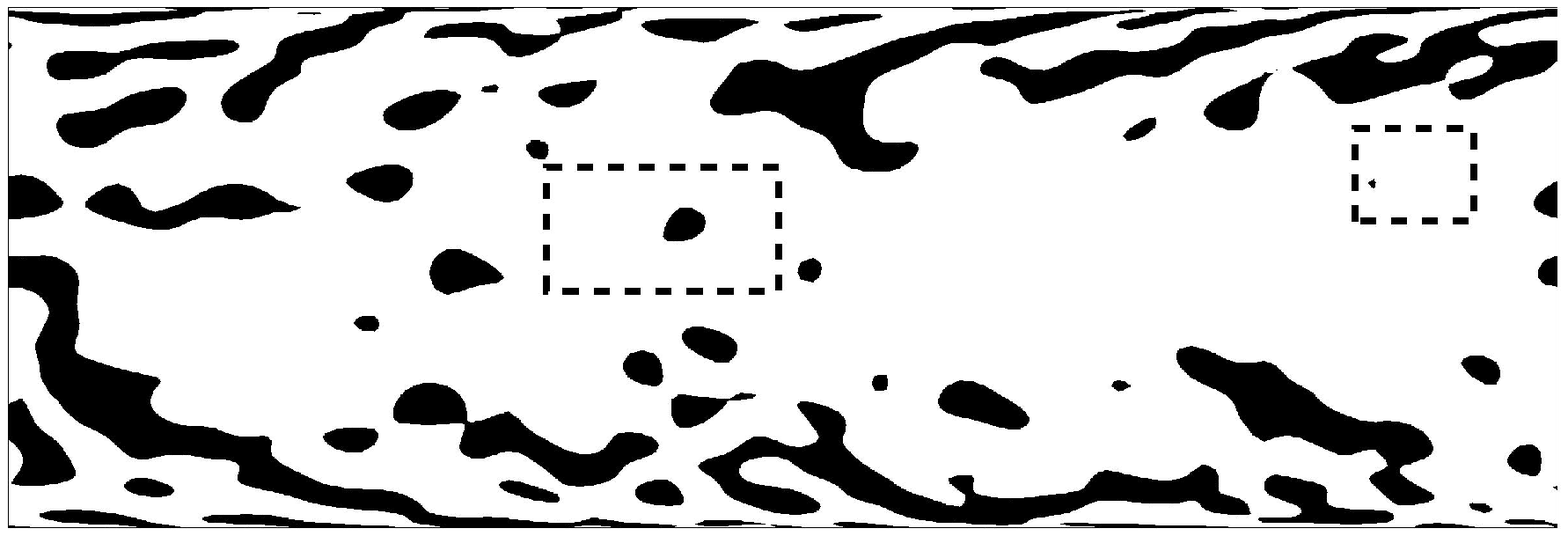}
\includegraphics[width=0.33\textwidth]{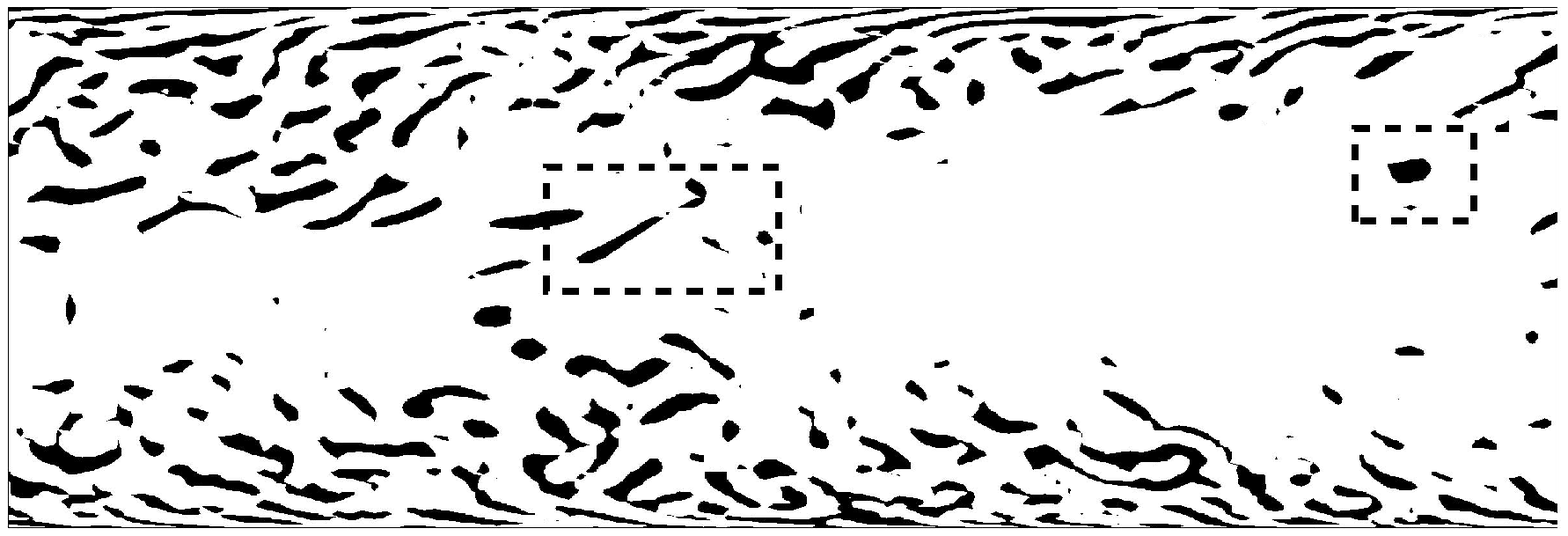}
}
\centerline{\it d)\hspace{0.3\textwidth} e) \hspace{0.3\textwidth} f)}
\caption{Streaks shown by the cut-off of the longitudinal velocity at $z^+=5.6,$ {\it a, b, c,} and streamwise vorticity in the wall-normal main-flow-parallel ($xz$) plane, {\it d,e,f.} Full flow, {\it a} and {\it d};   master-mode-set, {\it b} and {\it e};  remaining modes, {\it c} and {\it f.} For $\Rey_\tau=360,$  $L_x\times L_y=6\times3,$ and minimal master-mode set ($N_m=29716$). \label{Streaks}}
\end{figure}

Figure~\ref{Streaks} shows the visualisation of the near-wall  streaks in a wall-parallel ($xy$) plane $z^+=5.6$ and the visualisation of the streamwise vorticity in the wall-normal main-flow-parallel ($xz$) plane. It is clearly seen that all the streaks are contained in the minimal master-mode set. The same is definitely true for the large-scale vorticity structures, while some of the smaller vorticity structures, as indicated by the dashed line boxes, are actually found in the slave-set but not in the master-mode set. It would be interesting to follow the time history of these small structures in order to establish their dynamical significance; however, this is beyond the scope of the present study. The comparisons made so far show only that the minimal master-mode-set captures most of the structures.

\section{Master-mode database and an example of its use}\label{MMdatabase}
\subsection{Online database}

The properties of the master-mode-set, namely, its relatively small size and the possibility of recovering all other modes if necessary, make the master-mode-set suitable for creating a time-history database of a turbulent flow. For such a database a compromise should be made between the number of modes recorded and the recovery rate $\alpha$ (see~(\ref{alpha})). Increasing the number of modes above the minimal master-mode set size also improves the accuracy with which the master-mode set approximates the flow. In the course of the present study two master-mode-set databases were created, both for the channel flow at $\Rey_\tau=360.$ The first database contains a time history of the master-mode-set with $N_m=51508$ modes, which is about twice the size of the minimal master-mode set, for the box size $L_x\times L_y=6\times3$ and the duration $T=40.$ The convergence rate given by (\ref{alpha}) for this database is $\alpha\approx 10.$ Therefore, achieving 1\% accuracy of the recovered solution requires the slave mode calculation time $\Delta T=\alpha^{-1}\ln 0.01\approx0.4.$ The total size of the database is 112.7Gb. A fully resolved database would require about 2.6Tb of storage, other things being equal. The second database is for the box size $L_x\times L_y=14\times9$ with $N_m=2170044,$ which is about five times the size of the minimal master-mode-set, $T=15,$ and $\alpha\approx 160,$ which gives $\Delta T\approx0.03.$ It uses 1.78Tb of disk storage. A fully resolved database would require about 15.2Tb of storage, other things being equal.

\begin{figure}
\centerline{\raisebox{0.25\textwidth}{$R_{ii}$}
\includegraphics[width=0.7\textwidth]{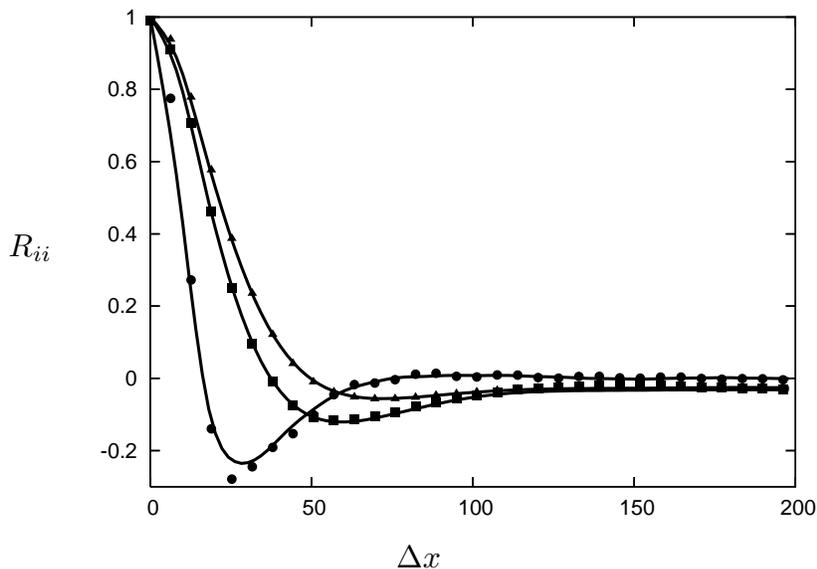}}
\centerline{$\Delta x$}
\caption{Spanwise ($\Delta x=0$) two-point correlations: curves, full flow as calculated by \cite{HuMorfeySandham06}, $L_x\times L_y=12\times6$; symbols, master-mode-set database with $N_{mm}=2170044$ and $L_x\times L_y=14\times9$; longitudinal velocity correlation, $R_{uu},$ $\blacksquare;$ spanwise velocity correlation, $R_{vv},$ $\blacktriangle;$ wall-normal velocity correlation, $R_{ww}$, $\bullet.$ For $\Rey_\tau=360$ and $z^+=10.$ \label{SpanwiseDBCor}} 
\end{figure}

As an alternative to recovering the slave modes one could simply store a set of instantaneous flow fields of the original calculations with the time interval $\Delta T,$ and recover the time history by running the full Navier-Stokes calculation using the nearest stored flow field as the initial condition. The calculations needed for recovering the slave modes are of a similar volume to full flow calculations, so, the CPU time required would be of the same order of magnitude in both approaches. For the second approach one would need to store $T/\Delta T$ snapshots, which would require only about 3.2Gb storage for the first database and 270Gb storage for the second database. However, in order to reproduce exactly the same flow by running a full Navier-Stokes calculation starting from the stored snapshot one would need to use a computer of the same architecture as that on which the original calculation was performed, as otherwise rounding errors would lead to the deviation of the original and reproduced solution due to its instability. In contrast, recovery of the slave modes is a stable procedure.
The master-mode database has the advantage of the possibility to use it as an approximate time history description of the flow without doing any recovery at all. The increase in size of the databases as compared to the minimal master-mode set size ensures a better approximation of the full velocity field. Figure~\ref{SpanwiseDBCor} shows the comparison for the spanwise two-point correlations calculated by \cite{HuMorfeySandham06} and those obtained from the larger master-mode-set database. Other quantities give at least as good an agreement. Moreover, the increased number of modes in the larger database removed the near-wall discrepancy illustrated by figure~\ref{Budgets}. Overall, the larger database gives quite a reasonable approximation to the full flow.

The relatively small size of the master-mode databases allows opening a free online access. The databases are currently at http://www.dnsdata.afm.ses.soton.ac.uk/.  The system allows the user to upload a Fortran 90 code, compile and run it on the server with access to the data through a subroutine library. The user can then download the output of the code.
This provides additional savings of the network traffic and appears to be more convenient than downloading the entire database to the remote computer.

\subsection{Travelling waves}

Using a master-mode-set database is most advantageous when an analysis of the same data needs to be repeated many times, as when searching for rare events and repeatedly adjusting the search criterion. Looking for travelling waves in a developed turbulent flow gives an example of such a procedure. Travelling wave solutions attracted recently much attention. For plane channel flow they were found by \cite{ItanoToh01} and \cite{Waleffe01,Waleffe03}, while the most recent attempt to find travelling waves in the velocity fields produced by \DNS\ was performed by \cite{KerswellTutty07}, where further references can be found. Note that as the Reynolds number and the size of the computational domain increase the likelihood of finding a travelling wave occupying the entire computational domain decreases. Therefore, one has to look for travelling waves in sub-volumes of the domain.

\begin{figure}
\centerline{
\includegraphics[width=0.33\textwidth]{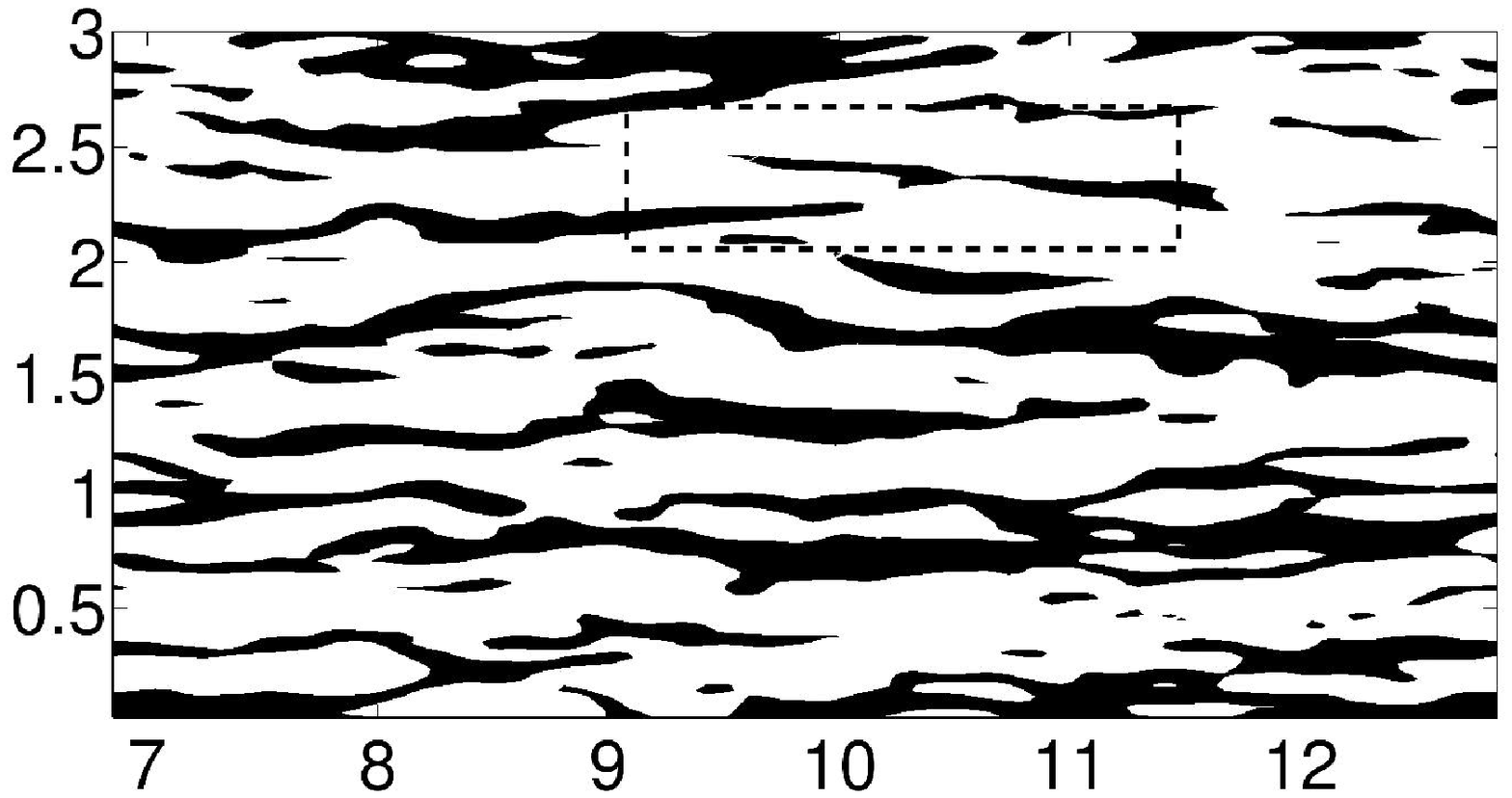}
\includegraphics[width=0.33\textwidth]{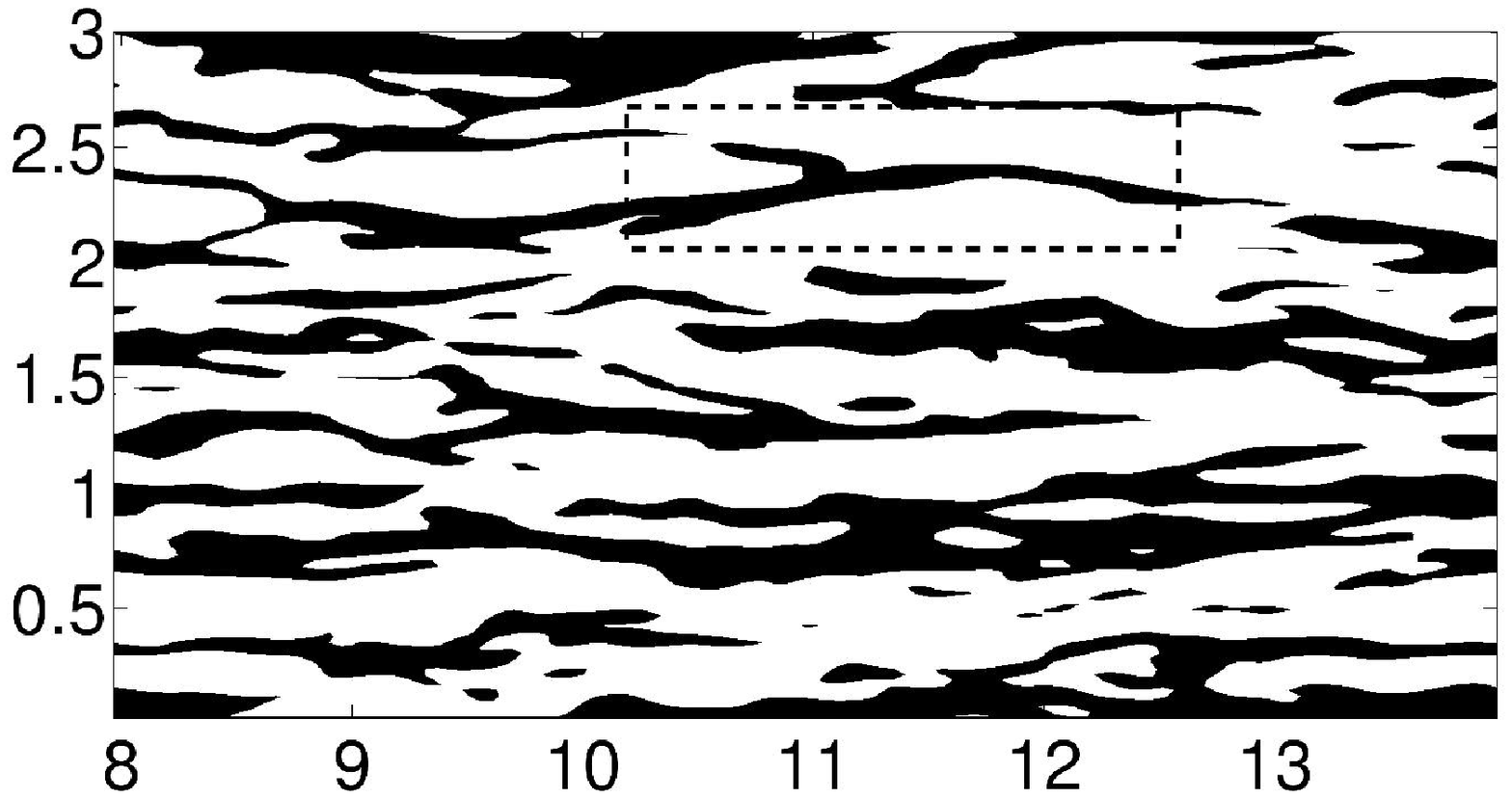}
\includegraphics[width=0.33\textwidth]{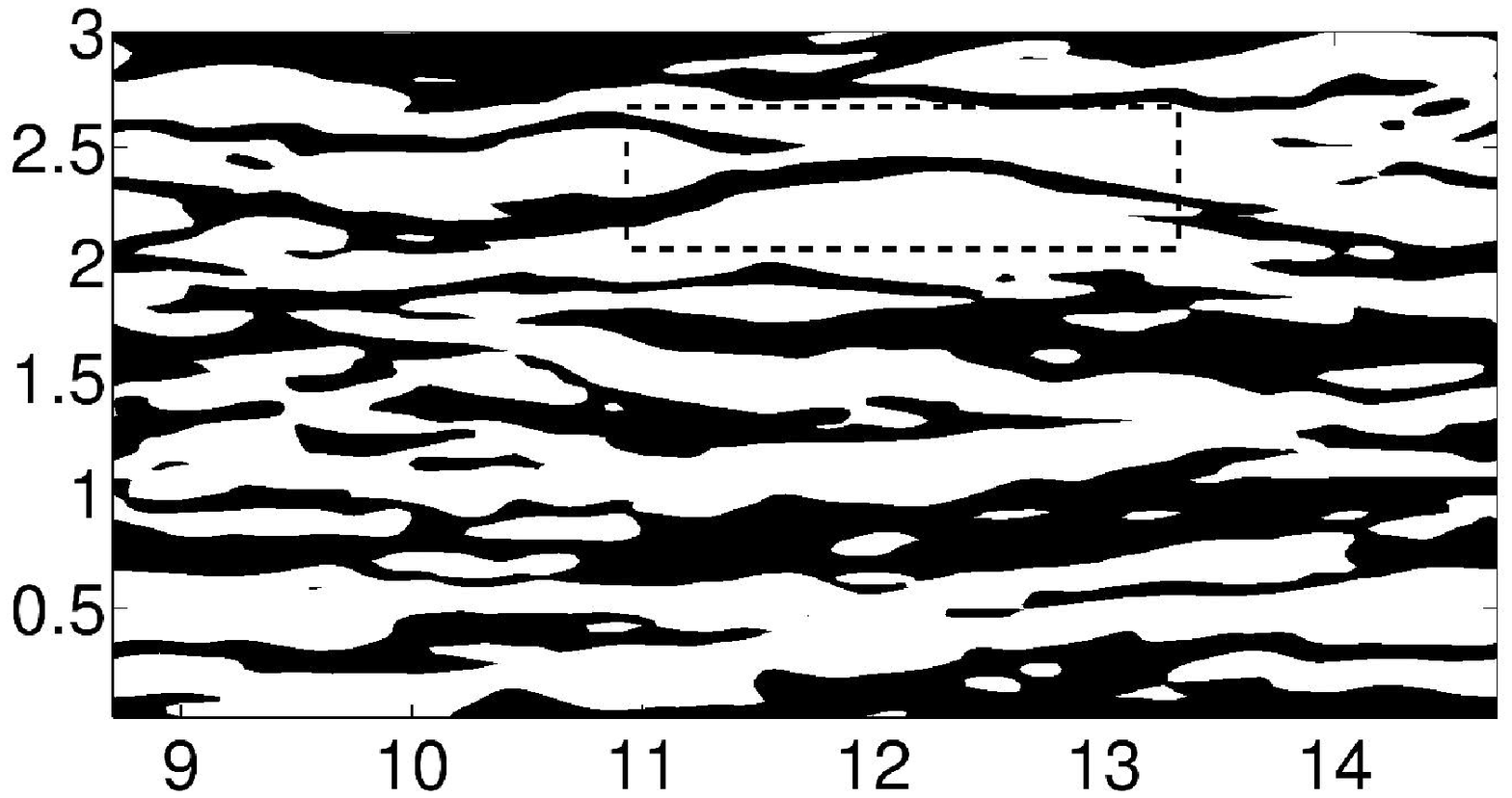}
}
\centerline{$t=13.625$\hspace{0.25\textwidth} $t=13.7$ \hspace{0.25\textwidth} $t=13.75$}

\

\centerline{
\includegraphics[width=0.33\textwidth]{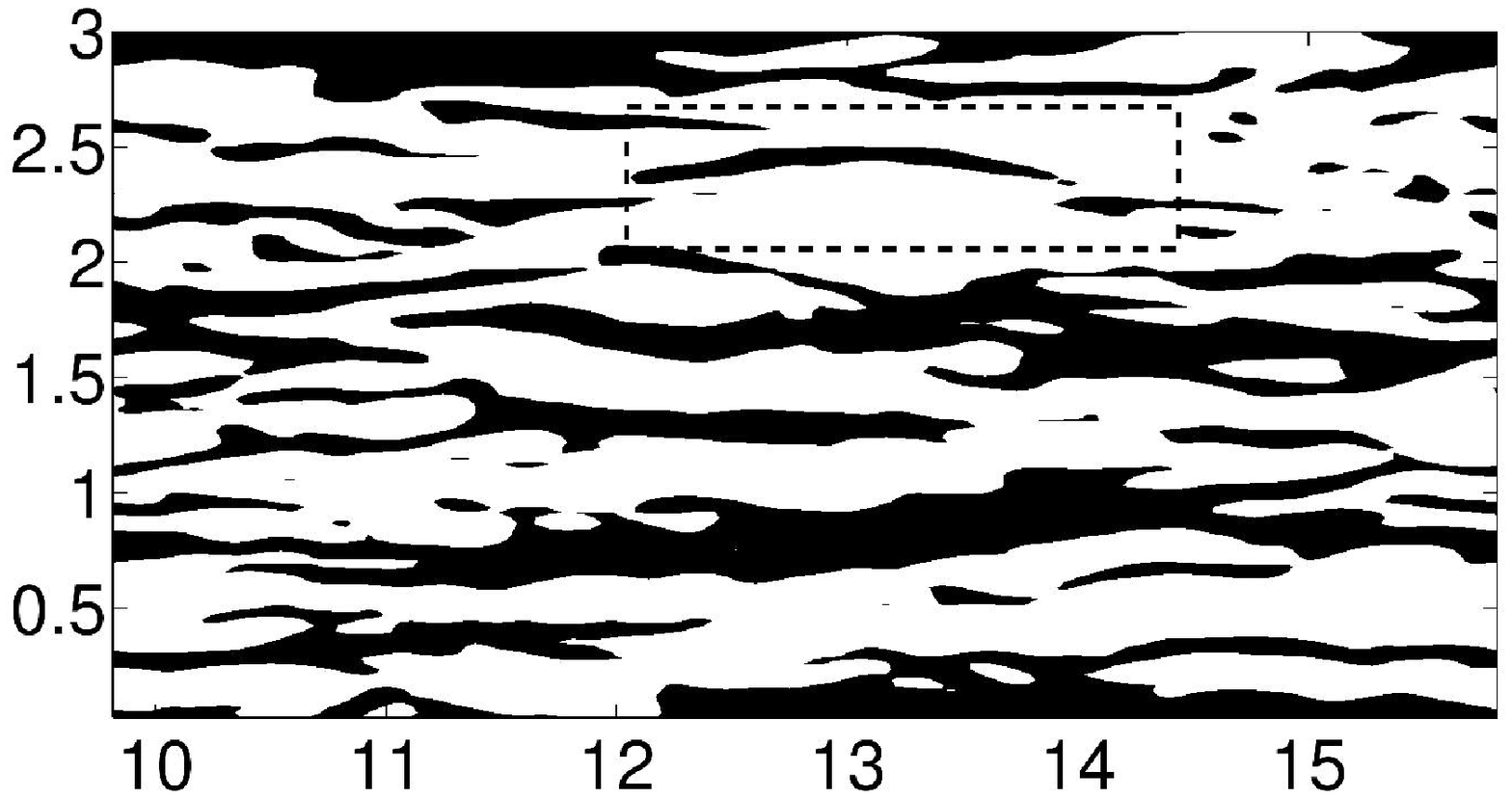}
\includegraphics[width=0.33\textwidth]{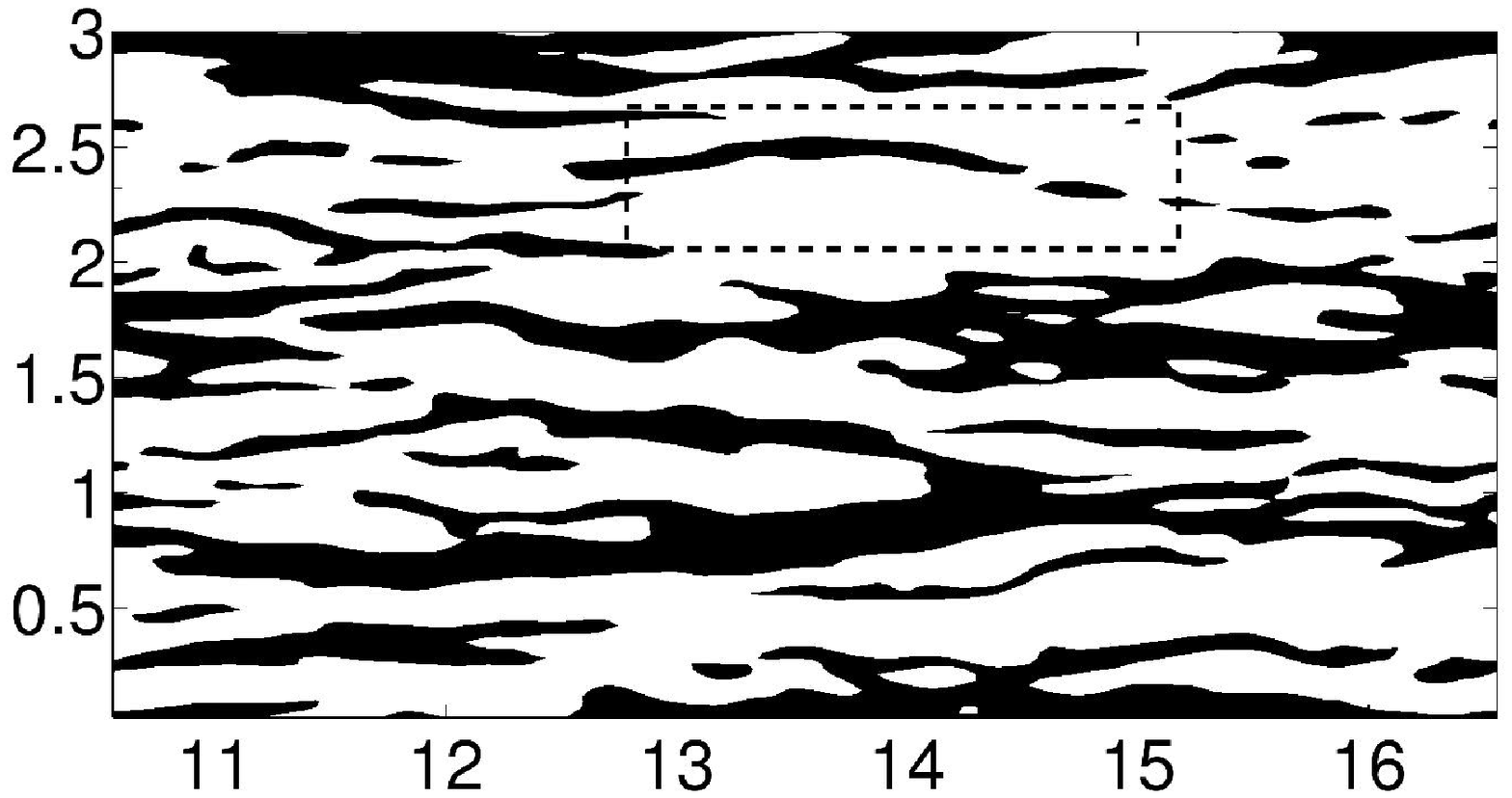}
\includegraphics[width=0.33\textwidth]{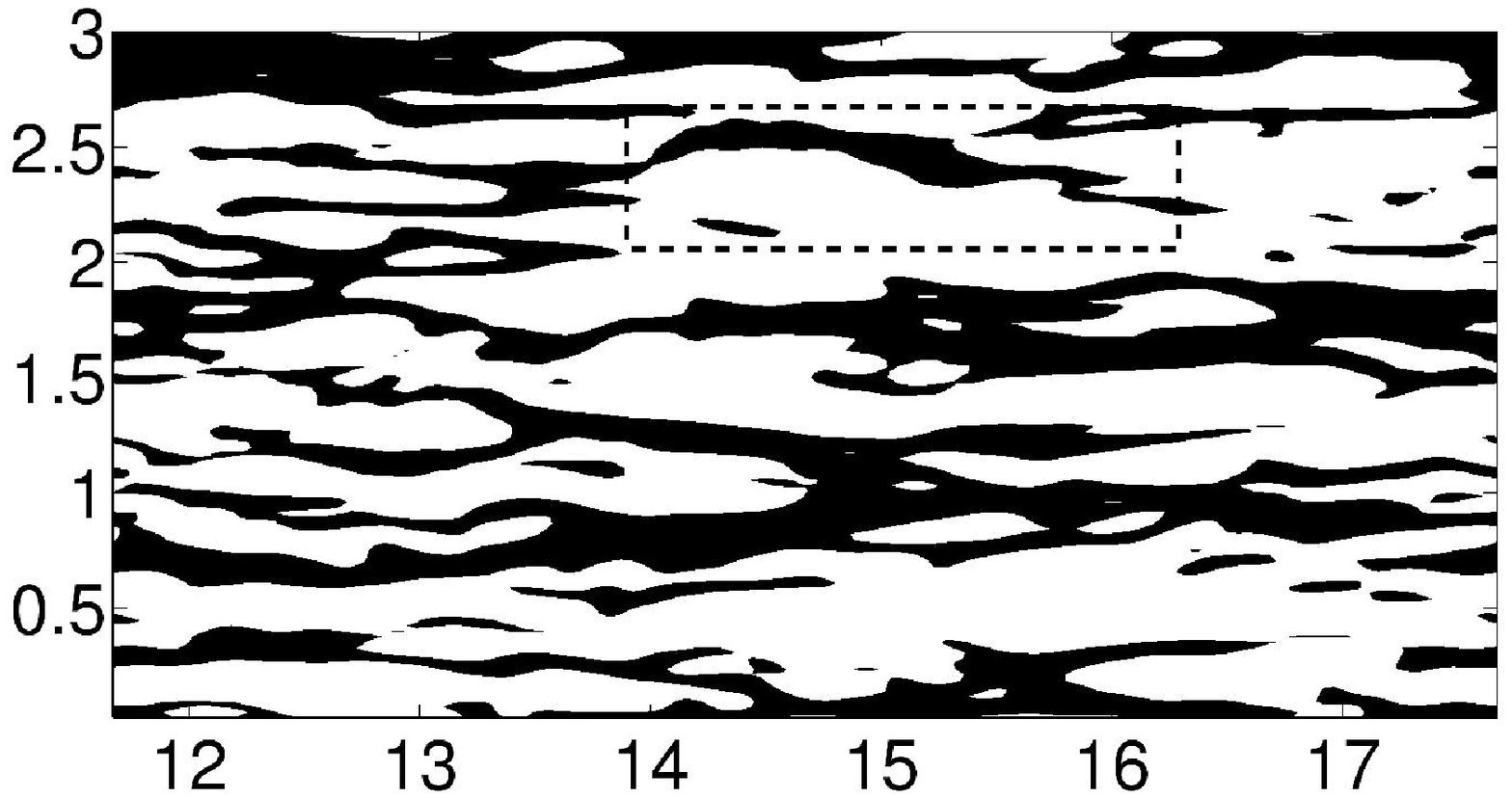}
}

\centerline{$t=13.825$\hspace{0.25\textwidth} $t=13.875$ \hspace{0.25\textwidth} $t=13.95$}
\caption{Travelling-wave-like object shown by the longitudinal velocity fluctuation at $z^+=6.9.$ $\Rey_\tau=360,$  $L_x\times L_y=6\times3,$ $N_{mm}=51508.$\label{TW}}
\end{figure}

A travelling wave solution has the form $\bu(x,y,z,t)=\bU(x-ct,y,z),$ where $c$ is the propagation speed of the travelling wave. For such solutions 
$$\pdd\bu{t}+c\pdd\bu{x}=0.$$ Accordingly, one can assume that a volume $W$ is occupied by a travelling wave if
$$I=\min_c\iiint_W\left|\pdd\bu{t}+c\pdd\bu{x}\right|^2dx\,dy\,dz$$
is small. To make this idea practical one needs, however, to define exactly what is meant here by small, that is to find a proper scale for $I.$ Also, the value of $c$ minimising this integral depends on $W.$ The procedure for looking for a travelling wave consists then in specifying a set of volumes $W$ and the scaling for $I,$ searching over a database for a minimum of the renormalized $I,$ looking at the outcome, refining intuitively both the volumes and the scaling and repeating the process again. This analysis was performed using the smaller database. Eventually, the scale for $I$ was chosen to be 
$$R=\iiint_W\left|\pdd\bu{t}\right|^2dx\,dy\,dz,$$
and the entire calculation domain $L_x\times L_y\times L_z=6\times3\times2$ was divided into rectangular boxes of the size $\Delta x\times \Delta y\times \Delta z =1.5\times0.15\times0.1.$ The minimum of $I/R$ was attained at $t=13.75$ in the volume $0<x<1.5,$ $2.4<y<2.55$, $-1<z<-0.9.$ Figure~\ref{TW} shows the visualisation of the longitudinal velocity at several consecutive times, with the area in question marked by a dashed rectangle. Note that the visual propagation speed in this figure is about $10,$ which is greater than the mean flow velocity at this distance to the wall (somewhat less than 7) but which is smaller than the velocity $c\approx14.8$ at which the minimum of $I/R$ was attained (the dashed rectangle moves at velocity $c$ in figure~\ref{TW}). By the visual propagation speed we understand the propagation speed of the visually identified point corresponding to the top of the dark-shaded arc in the middle of the dashed rectangle in the figures for $t=13.7-13.875.$ It is interesting to note that these velocities expressed in wall units (about $10$ and $15$) are respectively close to the propagation speed of velocity and vorticity fluctuations and the propagation speed of pressure fluctuations as identified by~\cite{KimHussain93}. Determining whether the object identified here is in fact related to travelling waves or whether it is a certain combination of generic near-wall structures is outside the scope of the present paper. The observations made here confirm the efficiency of the master-mode-set database as a tool for finding rare events in turbulent flow. The easy online access to the database gives an opportunity for any researcher to further investigate the object described in this section.

\section{Conclusions}

Master-modes have a wide scope and have not been previously studied for a three-dimensional turbulence in a channel flow. For this reason each of the results described in the present paper may be extended and deepened. One can now only summarise the outcomes of the present work:

\begin{itemize}
\item Master-mode sets exist for three-dimensional turbulent flow.
\item The minimal size of a master-mode set appears to be proportional to the product of the volume of the computational domain and the $9/4$ power of the Reynolds number $\Rey$ (based on the maximal velocity) for large $\Rey.$ A quantitative estimate may be inferred from figure~\ref{Dimension}. In a rather approximate way the size of the minimal master-mode-set can be estimated as about 2\% of the minimal number of modes needed to resolve the flow. These estimates may be dependent on the basis used in the calculations; in the present calculations it was Chebyshev polynomials in the wall-normal direction and Fourier expansions in the wall-parallel directions.
\item For channel flow the rate of convergence at which the $L_2$ norm of the difference between the master solution and the slave solution tends to zero can be estimated as $30(N/V)^{2/3}/\Rey_{\tau}$ for large $N,$ where $N$ is the master-mode set size (that is half of the number of independent real-valued parameters needed to identify all the mode amplitudes in the master-mode set), $V$ is the volume of the computational domain, and $\Rey_{\tau}$ is the Reynolds number based on the friction velocity and the channel half-width.
\item The velocity field corresponding to the minimal master-mode set is a good approximation for mean velocity and some other mean characteristics. However, the minimal master-mode set based on Chebyshev polynomials in the wall-normal direction produces large deviations of turbulence kinetic energy budget components from their exact values in the close vicinity of the wall where Chebyshev polynomials over-resolve the flow. Also, there are significant deviations for some of the two-point correlations. Therefore, it is reasonable to store in a database a master-mode set of a somewhat greater size than minimal.
\item When the flow field is represented as a sum of a master-mode set and the rest of the modes, the near-wall streaks turn out to be contained in the master-mode part. The results for the distribution of the longitudinal vorticity are less certain, with some small-scale structures sometimes observed in the rest of the modes.
\item A database containing the time history of a master-mode set is found to be an efficient tool for investigating rare events. In particular, a travelling-wave-like object was identified on the basis of the analysis of the database.
\item Two master-mode-set databases of the time history of a turbulent channel flow are available online at   http://www.dnsdata.afm.ses.soton.ac.uk/. The code uploaded by a user can be run on the server with an access to the data. \end{itemize}

\section{Acknowledgments}

Profs. N.\,Sandham, P.\,Spalart, and A.\,Tsinober made useful comments on the first draft of this paper. 
The work was supported by EPSRC under grant number GR/S67029. Partial support from the UKTC (EPSRC EP/D044073/1) and the Turbulence Platform (EPSRC GR/S82947/01) is also acknowledged. 

\bibliographystyle{jfm}
\bibliography{ChernyshenkoMasterModesArxiv}
\end{document}